\documentclass[preprintnumbers,amsmath,amssymb,superscriptaddress,twocolumn,showpacs]{revtex4-2}
\usepackage{graphicx}
\usepackage{dcolumn}
\usepackage{bm}
\usepackage{physics}
\usepackage[caption=false]{subfig}

\newcommand{\be}{\begin{equation}}
\newcommand{\ee}{\end{equation}}
\newcommand{\bea}{\begin{eqnarray}}
\newcommand{\eea}{\end{eqnarray}}

\begin{document}

\title[Planar Magnetic Paul Traps for Ferromagnetic Particles]{Planar Magnetic Paul Traps for Ferromagnetic Particles}
%
%

\author{M. Perdriat}
\author{C. Pellet-Mary}
\author{T. Copie}
\author{G. H\'etet} 

\affiliation{$^{1}$ Laboratoire De Physique de l'\'Ecole Normale Sup\'erieure, \'Ecole Normale Sup\'erieure, PSL Research University, CNRS, Sorbonne Universit\'e, Universit\'e Paris Cit\'e , 24 rue Lhomond, 75231 Paris Cedex 05, France.}

\date{\today}

\begin{abstract}
We present a study on the trapping of hard ferromagnetic particles using alternating magnetic fields, with a focus on planar trap geometries.
First, we realize and characterize a magnetic Paul trap design for millimeter-size magnets based on a rotating magnetic potential. Employing a physically rotating platform with two pairs of permanent magnets with opposite poles, we show stable trapping of hard ferromagnets a centimeter above the trap and demonstrate that the particle shape plays a critical role in the trapping.  Finally, we propose a chip trap design that will open a path to studies of gyromagnetic effects with ferromagnetic micro-particles.
\end{abstract}

\maketitle

In the 50's, Wolfgang Paul proposed to employ alternating electric fields to solve the conundrum imposed by Maxwell's equations \citep{WPaul_review} which prohibits levitation of charged particles using static electric fields.
Although the electric field is zero on average, the slight particle displacement during one period of the electric field alternation can lead to an efficient dynamical stabilization of the particle. 
Since then, Paul traps found countless applications such as in mass spectrometry and RF spectroscopy \cite{WELLING199895}, ground state cooling of atoms \citep{Monroe}, quantum computing \cite{Cirac} or more recently to quantum engineering of levitating particles \citep{Perdriat, Gonzalez}.
Trapping magnets is also an important endeavor that gained interest lately.
A particularly intriguing direction is the search for atomic-like effects on magnet motion, stemming from the spin degree of freedom \cite{Kimball, Rusconi, Kustura}. Observing such effects is under reach using trapped nano-ferromagnets or particles containing a large number of polarized spins and could lead to several applications in gyroscopy, magnetometry \cite{Vinante}, spin-mechanics \cite{Huillery, Gieseler3}, or in fundamental tests of quantum mechanics \cite{Timberlake}. 


There exists several magnetic levitation protocols for ferromagnets. For instance, adding angular momentum to a magnet on top of toroidal magnetic field can provide stable trapping. The mechanisms behind this so-called Levitron were attributed to the combined action of the gyroscopic stability and magnet's precession \cite{levitron}, both of which act to continuously align the magnet
precession axis to the local magnetic field direction. 
Another widely used method is electromagnetic suspension (EMS) \cite{maglev}, 
which uses servo-loops to counteract deviations of the particle motion away from a desired working point. 
A last example that shows efficient magnet levitation employs the diamagnetism of superconductors \cite{Druge_2014, NEMOSHKALENKO1990481, Gieseler3, Hull}, with large reported mechanical quality factors \cite{Gieseler3, Vinante3} and foreseeably high control in the engineering of the magnet quantum motion. 

\begin{figure}[ht!]
\centerline{\scalebox{0.16}{\includegraphics{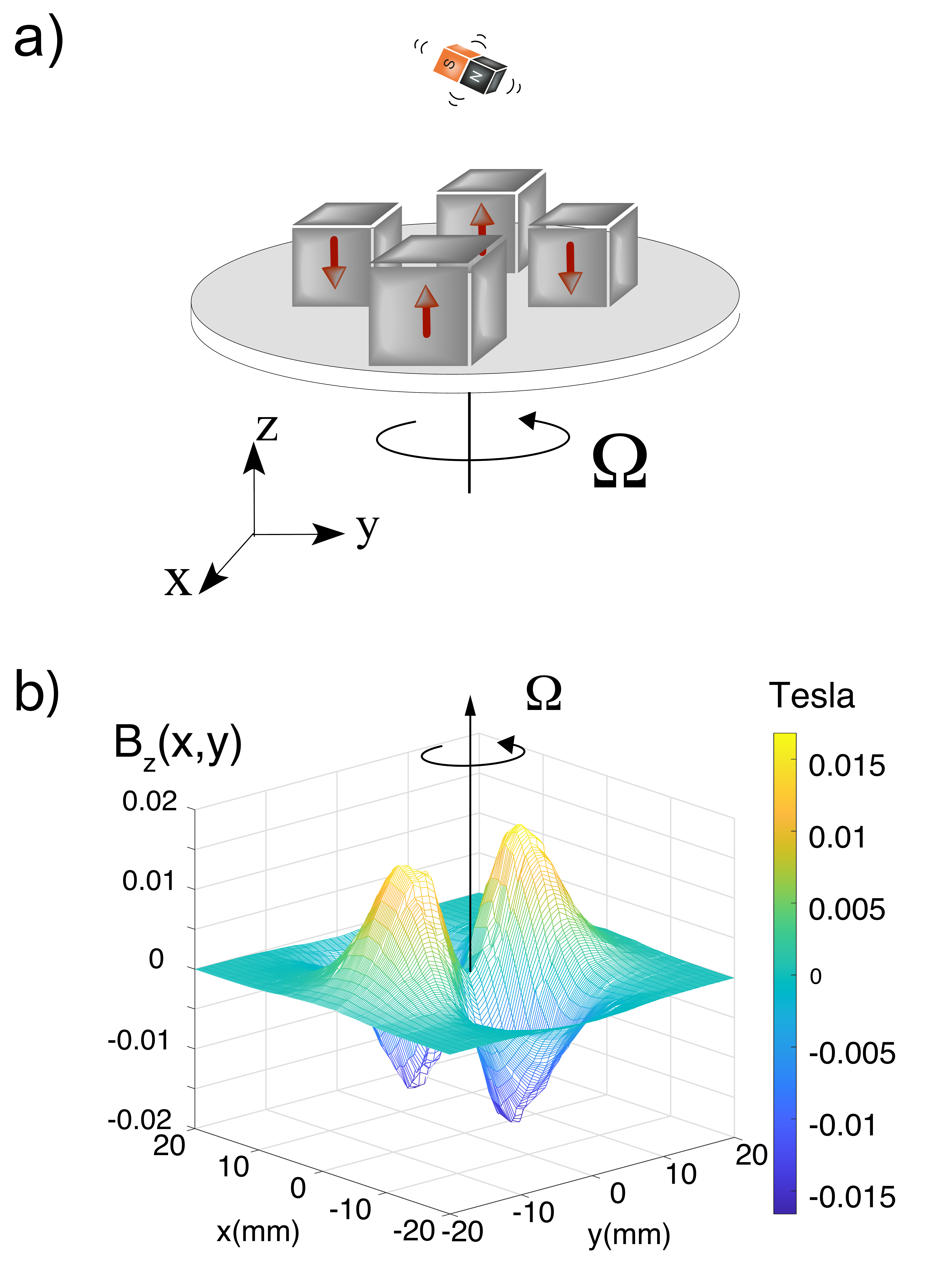}}}
\caption{a) Sketch of the magnetic Paul trap. b) Numerical simulations showing the projection of the magnetic field vector on the $z$ axis as a function of $x$ and $y$, at a distance $z=1$ cm from the trap.}
\label{Setup}
\end{figure}

Following the same logic as for charged particles, a trap that uses alternating magnetic fields may also offer strong harmonic confinement of magnets together with room temperature operation. A major difference between magnetic Paul traps (MPT) and electric Paul traps is that the magnetic energy depends on the ferromagnet orientation and thus the angular dynamics must be also be controlled for stable levitation. 
A strong enough homogeneous magnetic field can nevertheless fix the direction of the magnet dipole moment without exerting any force on the magnet. The method was in fact already realized \cite{Sackett} shortly after having been employed for trapping neutral atoms in the early days of Bose-Einstein condensation \cite{CornelNeutral}. The authors of Ref.~\cite{Sackett} used a combination of permanent magnets and time varying currents in Helmholtz coils to enable ponderomotive confinement. Similar traps for large particles were then soon implemented \cite{BASSANI, BASSANI2, BASSANI3}. 
In the present article, we demonstrate and characterize a magnetic Paul trap using alternating magnetic fields coming from physically rotating permanent magnets in a planar geometry. We then propose a planar on-chip design that will enable trapping of micro-metric particles, as well as enabling straightforward optical and inductive detection of their motion.


We start by showing the trapping of a ferromagnet for all 6 degrees of freedom, namely the 3 center of mass modes and the 3 librational modes, in a room temperature table-top set-up. 
Instead of the stationary-wave geometry typically employed in electric Paul traps, here we use the running-wave version, where the field curvatures in the $(x,y)$ plane rotate over time. We rotate magnetic fields using a physically rotating platform holding rigidly fixed permanent magnets to produce a ponderomotive potential in the $(x,y)$ plane. 

Figure \ref{Setup}-a) is a depiction of the platform that we employ. Opposite pairs of 5 mm square magnets are oriented so that their magnetic moments point to the same direction. The magnets are glued on the periphery of a copper baseplate. The distance between the center of the opposite magnet pairs is 20 mm. 
The assembly is attached to the spinning arm of the motor from a mechanical chopper. The motor can make this system rotate at a maximum frequency of 100 Hz, thus generating fast rotating magnetic fields along the $x,y$ directions. The $z$ direction can also be confined when the particle shape is taken into account, as will be explained later. 

Consider a particle of mass $m$ which can move in a plane subjected to a saddle potential rotating at the frequency $\Omega/2\pi$. We designate by $(x,y)$ the spatial coordinate of the particle in the laboratory-fixed frame and $(X,Y)$ the coordinate in the rotating saddle frame. Adding an extra homogeneous magnetic field along $z$ fixes the particle angle along the $z$ direction. As shown in the Supplemental Material (SM), because of the typically large moment of inertia of the particles we trap, only a moderate homogeneous B field (in the mT range) is enough to ensure that the particle magnetic moment direction is not being defined by the B fields from the rotating magnets.  

Figure \ref{Setup}-b) is the result of numerical simulations of the static magnetic field component along the $z$ direction as a function of $x$ and $y$ at a distance $z=1$ cm from the four-magnets trap. The magnetic field reaches about 0.1 T at maximum, and features a hyperbolic paraboloid shape close to $(x,y)=(0,0)$.
In the rotating frame, the magnetic field projection along $z$ around the saddle point reads : 
\begin{align}
B_z(X,Y)=\frac{1}{2}B_z''(z)\left(X^2-Y^2\right),
\end{align}
where $B_z''(z)$ is the curvature of the magnetic field component along $z$.
In this frame, two forces must be added: the centrifugal force $\mathbf{F}_{\rm cen}=m\Omega^2 (X\mathbf{e}_X+Y \mathbf{e}_Y)$ and the Coriolis force $\mathbf{F}_{\rm Cor}=-2m(\Omega \mathbf{e}_Z) \times (\dot{X}\mathbf{e}_X+\dot{Y} \mathbf{e}_Y) $. The confinement along $x$ and $y$ comes from the competition between $\mathbf{F}_{\rm cen}$ and $\mathbf{F}_{\rm Cor}$. In the laboratory-fixed frame, the magnetic field is time-dependent and reads:
\begin{align}
\begin{split}
    B_{z}(\mathbf{r},t)&=\frac{B_z''(z)}{2}\left((x^2-y^2)\cos{(2\Omega t)}-2xy\sin{(2\Omega t)} \right)
\end{split}.
\end{align} 

\begin{figure*}[ht!]
\centerline{\scalebox{0.35}{\includegraphics{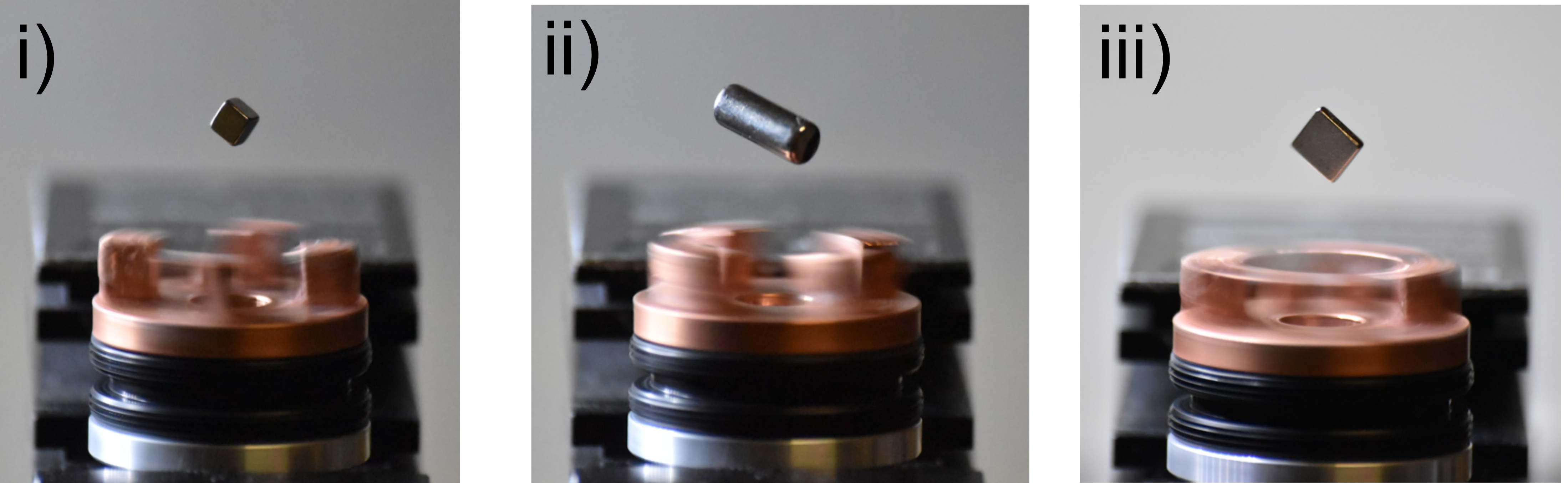}}}
\caption{Photos showing three magnets levitating above the rotating magnetic saddle: i) a 3 mm cube, ii) a 10 mm long cylinder with a 5 mm diameter and iii) a parallelepiped with short side-length of 1~mm and a 5 mm $\times$ 5 mm, flat square magnet. In these experiments, the magnets do not rotate over time in any direction (see movie in the Supplementary materials).}
\label{Pictures}
\end{figure*}

The coupled set of equations of motions for the center of mass coordinates subjected to $B_{z}(\mathbf{r},t)$ can be solved straightforwardly by averaging over one period of the trap oscillation (see SM, section 1).
This procedure leads to ponderomotive confinements both in the $x$ and $y$ directions at trapping frequencies that are given by 
\begin{align}
\frac{\omega_{x,y}}{2\pi}&={\frac{ B_{\rm sat} |B''_z(z)|}{\mu_0 \pi \rho_m \Omega}}.
\end{align}
Here, $\rho_m$ is the magnet density and $B_{\rm sat}\approx 1$ T is the magnetization at saturation of the magnet.
Using $\Omega/2\pi\approx 100$ Hz, we obtain $\frac{\omega_{x,y}}{2\pi}$ in the Hz range when the magnet lies a centimeter above the trap.
Notably, in this geometry, the magnetic field is zero at the center $(x,y)=(0,0)$ at any position along $z$. If the particle was a point-like magnet, there would be no confinement in the $z$ direction.
A homogeneously magnetized particle whose form is invariant by $\pi/2$ rotations about an axis $\bm e_s$, will also not experience any force along $z$ if the symmetry axis $\bm e_s$ coincides with $\bm e_z$. 
Indeed, the total force exerted by the four magnets on the whole body cancels out at any point in time. 
However, any deviation of the particle symmetry axis from the $z$ direction imposed by the external homogeneous magnetic field may give rise to a net force along $z$ onto the particle. When averaged over a cycle of the trap rotation, the particle would feel a ponderomotive force that pushes it away from the $z=0$ point. 
Another situation which could give rise to such a force along $z$ is when the particle is asymmetrical. Taking a parallelepiped with magnetic moment along $z$, with a length $l$ along $x$ and a square cross-section with a side length $h<l$, one finds (See SM section II) that the time-dependent potential along the $z$ direction reads : 
\begin{align}
E_{\rm mag}(\mathbf{r},t) \approx \frac{1}{2}m \omega_z^2 \cos{(2\omega t)} z^2,
\end{align}
where 
\begin{align}
\omega_z&=\sqrt{\frac{B_{\rm sat}a_2}{12 \mu_0 \rho_m}\left(l^2-h^2\right)}. \label{Eq2}
\end{align}
Note that the radial potential is not impacted by this shape effect to first order. 
As expected the force is proportional to $a_2=\partial^2 B''_z(z)/ \partial^2 z$ and
the decrease of the magnetic curvature $B''_z(z)$ along $z$ gives rise to an outward force when averaged over one oscillation period. There will, in turn, be a local energy minimum thanks to gravity, thus confining the particle at a distance $d\approx$~1 cm from the trap (see SM section 1). 


Experimentally, we find that at frequencies $\Omega/2\pi$ in the 50 to 100 Hz range, millimeter scale magnets can stably levitate for extended periods of time (more than hours) at a distance $d\approx 15$ mm from the rotating baseplate where magnets are glued. 
Fig \ref{Pictures} i) to iii) show three photos of levitating magnets of different shapes and sizes. All three are neodymium iron boron magnets with a nickel coating. 
The three magnets are perfectly stable for all 6 degrees of freedom and over extended periods of time. 
A video of a levitating magnet can also be found in the supplemental materials. 
We found that placing magnets in a small water container above the trap and then removing the water using a syringe once they are stably trapped ensures efficient loading. The ease with which we can trap under water most likely stems from the increased capture volume as the damping is increased. 
After optimizing external homogeneous magnetic field directions and amplitudes using an extra permanent magnet, all of these magnets can be made stable angularly.
The homogeneous magnetic field can indeed change the orientation of the magnet without perturbing the particle center of mass, as long as the magnetic field gradients at the particle location are negligible compared to the effective time-averaged magnetic field gradients from the saddle.
We also found that the stator of the motor itself generates a strong enough static magnetic field that can also orient the magnet. 

\begin{figure*}[ht!]
\centerline{\scalebox{0.25}{\includegraphics{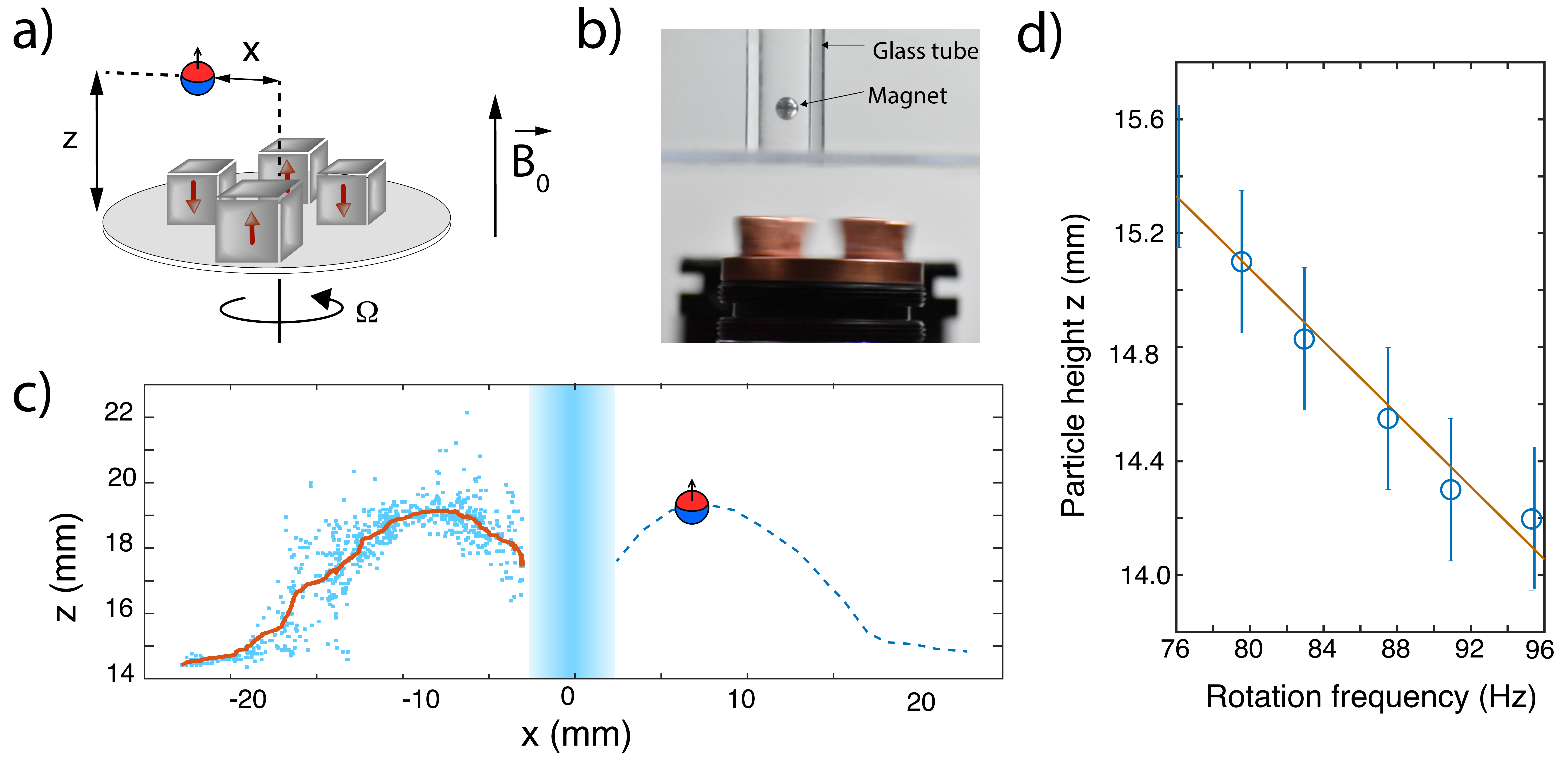}}}
\caption{a) Schematic showing the parametrization of the particle coordinates. b) Picture of the set-up showing the glass tube where the particle is enclosed. c) Measured position of the center of mass of a spherical 1 mm diameter magnet as a function of the distance from the platform rotation axis. The central blue area indicates the range of $x$ values where the particle sits at the bottom of the glass tube. d) Position of the center of mass of a square magnet along $z$ as a function of the rotation frequency $\Omega/2\pi$. }
\label{Setup2}
\end{figure*}

Numerical simulations (See SM, section 1-B) show good agreement with the experimental observations. In particular, the observed trapping height agrees with a model taking into account the particle shape. 
We also found further evidence that confinement along the $z$ direction arises because of the finite particle size by performing experiments with a millimeter-size spherical magnet. Confinement along $z$ was not observed for such particles, in agreement with Eq.~\ref{Eq2}, which predicts no confinement if $l\approx h$ in the trap center. Note that this is true only for small enough particles (See SM section I-B for a description of the limits of the model). In order to measure the $z$ dependent force when this particle is displaced from the trap center, we enclose the particle in a small glass tube and read-out its $z$ motion as the tube is displaced transversally. Fig. \ref{Setup2}-b) shows a picture of the set-up and 
Fig. \ref{Setup2}-a) is a schematic showing the employed parametrization. 
Fig. \ref{Setup2}-c) are data where the center of mass position $z$ is monitored as a function of $x$.
The particle experiences no outward force when the tube is positioned such that $|x|<2.5$~mm and remains at the bottom of the tube. As the tube is displaced, the particle is lifted up, with a maximum height at $|x|\approx 19$ mm, and then falls down again. This behavior is very well captured by numerical simulations where the outward force comes from the competition between ponderomotive force and gravity. Note also that, although extensive tests have been made, no trapping could be obtained after rotating the whole apparatus upside-down, which shows that the $z$ confinement is not solely due to the rotating magnetic field and that gravity plays a role.
One last check is the dependency of the particle position with $\Omega/2\pi$.
Fig.~\ref{Setup2}-d) shows the height $z$ of a trapped 2.5 mm side cubic magnet (Fig. \ref{Pictures}), as a function of $\Omega/2\pi$. $z$ is seen to increase as the trap rotation frequency is increased, as expected from the ponderomotive nature of the outward force competing with gravity.

\begin{figure}[ht!]
\centerline{\scalebox{0.44}{\includegraphics{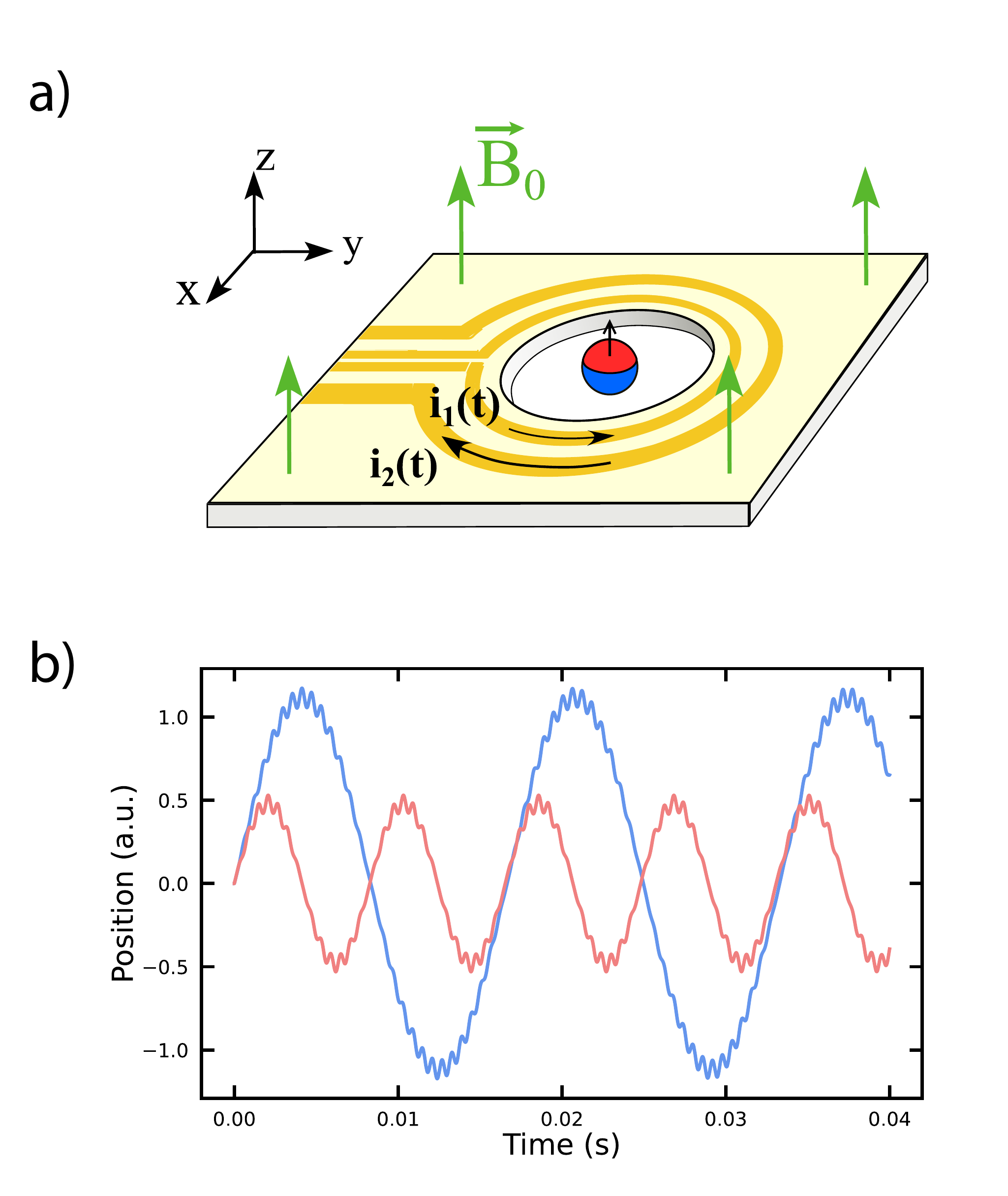}}}
\caption{a) Depiction of the current carrying trap design. b) Simulations showing the center of mass motion in the $x$ direction (blue curve) and the $z$ direction (red curve) for a trapping frequency $\Omega/2\pi=1.7$ kHz, an initial position $x_0=0$, $z_0=0$ and an initial velocity $\dot{x}_0 \ne 0$, $\dot{z}_0 \ne 0$.}
\label{Setup3}
\end{figure}

The off-centered position of the particle along $z$ is very convenient for the large optical access required for optical read-out of the motion. It will also facilitate the particle levitation in a vacuum chamber without having to integrate the rotating platform. Such planar traps are thus very attractive. 
As is the case with magnetic traps for neutral atoms or with ion traps, the most versatile way to trap particles while reaching high trapping frequencies together with open access is to employ chip-based designs consisting of current-carrying wires. 
Here, we propose the planar trap design depicted in Fig. \ref{Setup3}-a) where an alternating current runs through a micro-wire in an externally applied homogeneous field $\mathbf{B}_0=B_0\mathbf{e}_z$. 
The total applied magnetic field $\mathbf{B}_{\rm{tot}}(\mathbf{r},t)$ on the ferromagnet at the center of this ring is the sum of a homogeneous field $\mathbf{B}_0$ and an oscillating parabolic magnetic field $\mathbf{B}_1(\bm r,t)$ which reads 
\begin{eqnarray}\nonumber
    \mathbf{B}_1(\mathbf{r},t)&=& \frac{B_1''}{2} \Big[(z^2-\frac{1}{2}(x^2+y^2))\mathbf{e}_z \\
    &-&\left(xz\mathbf{e}_x+yz\mathbf{e}_y\right)\Big] \cos{(\Omega t)},
\end{eqnarray}
where $(\mathbf{e}_x,\mathbf{e}_y,\mathbf{e}_z)$ is the laboratory-fixed axis, $(x,y,z)$ are spatial coordinates, $B_1''$ is the curvature of $\mathbf{B}_1(\mathbf{r},t)$ and $\Omega/2\pi$ is the frequency of the magnetic field oscillation. 

We consider a hard ferromagnetic neodymium microsphere with a radius $a=1~\mu\rm{m}$. 
We use the $\bm{zyz}$ convention for the three Euler angles ($\alpha$, $\beta$, $\gamma$) that parametrize the orientation of the magnet as well as the body-fixed magnetic moment orientation $\bm{\mu}$ (see SM, section II). 
The norm of the magnetic moment is $\mu=B_{\rm{sat}}V/\mu_0$, where $V$ the volume of the ferromagnet and $B_{\rm{sat}}\approx 1.0~ \rm{T}$ is the magnetization at saturation. In the $(x,y,z)$ basis, the magnetic moment is expressed as 
\begin{eqnarray}\nonumber
\bm{\mu}&=&-\mu (-c_\alpha s_{\Tilde{\beta}} c_\gamma-s_\alpha s_\gamma) \mathbf{e}_x \\
&-& \mu(c_\alpha s_\gamma- s_\alpha s_{\Tilde{\beta}} c_\gamma)\mathbf{e}_y-\mu c_{\Tilde{\beta}} c_\gamma\mathbf{e}_z, 
\end{eqnarray}
where $c_i$ and $s_i$ designate the cosine and sine function with arguments $i$ and $\Tilde{\beta}$ is the shifted nutation angle defined as $\Tilde{\beta}=\beta-\pi/2$.
By performing a second order Taylor expansion of the microsphere magnetic energy $E_{\rm{mag}}=-\bm{\mu} \cdot \mathbf{B}_{\rm{tot}}(\mathbf{r},t)$ around the position $(\Tilde{\beta},\gamma, x ,y ,z)=0$, we obtain:
\begin{eqnarray} \nonumber
    E_{\rm{mag}}&=&\mu B_0 \left(\frac{\gamma^2}{2}+\frac{\Tilde{\beta}^2}{2} \right)\\
    &-&\frac{\mu B_1''}{2}\left(z^2-\frac{1}{2}(x^2+y^2)\right) \cos{(\Omega t)}.
\end{eqnarray}
Within the stability conditions (See SM, section 1-B), the alternating magnetic potential leads to a pseudo-harmonic confinement of the center of mass modes.
After averaging the potential over one period of the magnetic field oscillation, we obtain the secular frequencies:
\begin{align}
\omega_\gamma=\omega_{\Tilde{\beta}}=\sqrt{\frac{5}{2} \frac{B_0 B_{\rm{sat}}}{\mu_0 \rho_m a^2}}
\end{align}
and 
\begin{align}
\quad \Tilde{\omega}_z=2\Tilde{\omega}_x=2\Tilde{\omega}_y=\frac{\Omega}{2}\frac{|q_z|}{\sqrt{2}},
\end{align}
where 
$
q_z=-2q_x=-2q_y=2 B_1'' B_{\rm{sat}}/ (\mu_0\rho_m \Omega^2).$
Taking $q_z=0.2$, where the secular harmonic approximation is valid and the following experimentally accessible values : $B_0=10~\rm{mT}$,  $B_1'' = 10^5~\rm{T}.\rm{m}^{-2}$ and $\Omega = (2\pi) 2.0~\rm{kHz}$, we obtain $\omega_\gamma=\omega_{\Tilde{\beta}} = (2\pi) 300~\rm{kHz} $ and  $\Tilde{\omega}_z=2\Tilde{\omega}_x=2\Tilde{\omega}_y =(2\pi) 100~\rm{Hz}$. 
These trapping frequencies are on the order of the trapping frequencies that can be achieved using superconducting traps for micromagnets \cite{Gieseler3, Vinante3}, albeit here using a room temperature set-up.
Fig. \ref{Setup3}-b) shows the result of numerical simulations where the center of mass motion for the $x$ (blue curve) and $z$ (red curve) modes evolve as a function of time for a Paul trap frequency $\Omega/2\pi=1.7$ kHz.  We see that as the magnet is displaced away from the position $x,z=0$, a small high frequency oscillation appears after which the particle is brought back to $x,z=0$. This observation is in accordance with the expected principle of the ponderomotive confinement and in good agreement with the above analytical calculations. 

The shift of the center of mass position induced by gravity is $z_0=-g/\Tilde{\omega}_z^2 \approx -20 \mu \rm{m}$. Such a shift will generate micromotion in the $\mathbf{e}_z$ direction as well as a small coupling between the center of mass and the angular modes. We found that the gravity can however be compensated for by an external electric field if the particle is electrically charged or using a magnetic field gradient such that $\mathbf{B}_2=B_2'\left(z\mathbf{e}_z-x/2\mathbf{e}_x-y/2\mathbf{e}_y\right)$ with $B_2'=mg/\mu=8.0 \times 10^{-2} \rm{T}.\rm{m}^{-1}$. The coupling between the center of mass and the angular modes induced by this extra field can safely be neglected here (See SM, section 2). 

While it is straightforward to experimentally generate the magnetic field $\mathbf{B}_0$, the generation of an oscillating magnetic field with a curvature $B_1'' = 10^5~\rm{T}.\rm{m}^{-2}$ requires a carefully engineered trap. We propose to employ a high-purity gold double-loop micro-trap with radii $r_1=100~\mu \rm{m}$ and $r_2=200~\mu \rm{m}$ made by lithography, on a silicon chip with alternating currents $i_k(t) = i_k \cos{(\omega t)}$, $k=\{1,2 \}$ with $-2i_1=i_2$.  The condition $i_1/i_2=-r_1/r_2$ further suppresses the magnetic field at $\bm r =0$ while keeping its harmonic dependency. The magnetic field curvature reads  
 $B_1''=-\frac{9}{16} \mu_0 i_1/ r_1^3$.
This type of microchip can carry current of $10^5 \rm{A}/{cm}^2$ without experiencing too much heating \cite{PhysRevLett.84.4749, Feenstra}. With a $50~\mu \rm{m}$ large and $2~\mu \rm{m}$ thin gold layer, the current $i_1$ equals $0.1$ A which leads to the desired value $B_1'' \approx 10^5~\rm{T}.\rm{m}^{-2}$. The induced currents generated by one loop onto the other are found to be negligible (See SM, section 3). The Eddy currents created inside the levitating ferromagnet are also negligible compared to the gas dissipation process even under high vacuum (see SM, section 3). We thus expect no heating of the particle in this parameter regime.


The ability to control translational and rotational degrees of freedom of trapped particles with high precision has led to new opportunities for fundamental and applied research \cite{Gonzalez}. The high sensitivities of levitated objects to forces and torques in the motional ground state motivates research on tests of spin-mechanical coupling in the mesoscopic regime where gyromagnetic effects play a role.
It is arguably beneficial to enrich the set of tools available to release some constraints that may be present in some trapping systems. As an example, optical levitation is typically prone to particle heating \cite{Rahman, Neukirch, Hoang} and trapped charged particles to not lend themselves easily to approaching objects such as micro-fabricated optical components or other two-level systems \cite{Huillery} at distances below the micron. The presented magnetic Paul trap seems not to suffer from these problems and will thus enrich  this already multidisciplinary research field. 

\begin{acknowledgments}
We would like to thank fruitful discussions with R. Folman, D. Budker, J. Wei, R. Blatt, O. Romero-Isart and P. Huillery. GH acknowledges funding from the Ile-de-
France region in the framework of the DIM SIRTEQ. This project also received funding from the European Union's Horizon 2020 Research and Innovation Programme under Grant Agreement no. 731473 and 101017733. 
\end{acknowledgments}

\bibliography{MainMPT.bib}

\begin{thebibliography}{29}%
\makeatletter
\providecommand \@ifxundefined [1]{%
 \@ifx{#1\undefined}
}%
\providecommand \@ifnum [1]{%
 \ifnum #1\expandafter \@firstoftwo
 \else \expandafter \@secondoftwo
 \fi
}%
\providecommand \@ifx [1]{%
 \ifx #1\expandafter \@firstoftwo
 \else \expandafter \@secondoftwo
 \fi
}%
\providecommand \natexlab [1]{#1}%
\providecommand \enquote  [1]{``#1''}%
\providecommand \bibnamefont  [1]{#1}%
\providecommand \bibfnamefont [1]{#1}%
\providecommand \citenamefont [1]{#1}%
\providecommand \href@noop [0]{\@secondoftwo}%
\providecommand \href [0]{\begingroup \@sanitize@url \@href}%
\providecommand \@href[1]{\@@startlink{#1}\@@href}%
\providecommand \@@href[1]{\endgroup#1\@@endlink}%
\providecommand \@sanitize@url [0]{\catcode `\\12\catcode `\$12\catcode
  `\&12\catcode `\#12\catcode `\^12\catcode `\_12\catcode `\%12\relax}%
\providecommand \@@startlink[1]{}%
\providecommand \@@endlink[0]{}%
\providecommand \url  [0]{\begingroup\@sanitize@url \@url }%
\providecommand \@url [1]{\endgroup\@href {#1}{\urlprefix }}%
\providecommand \urlprefix  [0]{URL }%
\providecommand \Eprint [0]{\href }%
\providecommand \doibase [0]{https://doi.org/}%
\providecommand \selectlanguage [0]{\@gobble}%
\providecommand \bibinfo  [0]{\@secondoftwo}%
\providecommand \bibfield  [0]{\@secondoftwo}%
\providecommand \translation [1]{[#1]}%
\providecommand \BibitemOpen [0]{}%
\providecommand \bibitemStop [0]{}%
\providecommand \bibitemNoStop [0]{.\EOS\space}%
\providecommand \EOS [0]{\spacefactor3000\relax}%
\providecommand \BibitemShut  [1]{\csname bibitem#1\endcsname}%
\let\auto@bib@innerbib\@empty
\bibitem [{\citenamefont {Paul}(1990)}]{WPaul_review}%
  \BibitemOpen
  \bibfield  {author} {\bibinfo {author} {\bibfnamefont {W.}~\bibnamefont
  {Paul}},\ }\bibfield  {title} {\bibinfo {title} {Electromagnetic traps for
  charged and neutral particles},\ }\href
  {https://doi.org/10.1103/RevModPhys.62.531} {\bibfield  {journal} {\bibinfo
  {journal} {Rev. Mod. Phys.}\ }\textbf {\bibinfo {volume} {62}},\ \bibinfo
  {pages} {531} (\bibinfo {year} {1990})}\BibitemShut {NoStop}%
\bibitem [{\citenamefont {Welling}\ \emph {et~al.}(1998)\citenamefont
  {Welling}, \citenamefont {Schuessler}, \citenamefont {Thompson},\ and\
  \citenamefont {Walther}}]{WELLING199895}%
  \BibitemOpen
  \bibfield  {author} {\bibinfo {author} {\bibfnamefont {M.}~\bibnamefont
  {Welling}}, \bibinfo {author} {\bibfnamefont {H.}~\bibnamefont {Schuessler}},
  \bibinfo {author} {\bibfnamefont {R.}~\bibnamefont {Thompson}},\ and\
  \bibinfo {author} {\bibfnamefont {H.}~\bibnamefont {Walther}},\ }\bibfield
  {title} {\bibinfo {title} {Ion/molecule reactions, mass spectrometry and
  optical spectroscopy in a linear ion trap},\ }\href
  {https://doi.org/https://doi.org/10.1016/S0168-1176(97)00251-6} {\bibfield
  {journal} {\bibinfo  {journal} {International Journal of Mass Spectrometry
  and Ion Processes}\ }\textbf {\bibinfo {volume} {172}},\ \bibinfo {pages}
  {95} (\bibinfo {year} {1998})}\BibitemShut {NoStop}%
\bibitem [{\citenamefont {Monroe}\ \emph {et~al.}(1995)\citenamefont {Monroe},
  \citenamefont {Meekhof}, \citenamefont {King}, \citenamefont {Jefferts},
  \citenamefont {Itano}, \citenamefont {Wineland},\ and\ \citenamefont
  {Gould}}]{Monroe}%
  \BibitemOpen
  \bibfield  {author} {\bibinfo {author} {\bibfnamefont {C.}~\bibnamefont
  {Monroe}}, \bibinfo {author} {\bibfnamefont {D.~M.}\ \bibnamefont {Meekhof}},
  \bibinfo {author} {\bibfnamefont {B.~E.}\ \bibnamefont {King}}, \bibinfo
  {author} {\bibfnamefont {S.~R.}\ \bibnamefont {Jefferts}}, \bibinfo {author}
  {\bibfnamefont {W.~M.}\ \bibnamefont {Itano}}, \bibinfo {author}
  {\bibfnamefont {D.~J.}\ \bibnamefont {Wineland}},\ and\ \bibinfo {author}
  {\bibfnamefont {P.}~\bibnamefont {Gould}},\ }\bibfield  {title} {\bibinfo
  {title} {Resolved-sideband raman cooling of a bound atom to the 3d zero-point
  energy},\ }\href@noop {} {\bibfield  {journal} {\bibinfo  {journal} {Phys.
  Rev. Lett.}\ }\textbf {\bibinfo {volume} {75}},\ \bibinfo {pages} {4011}
  (\bibinfo {year} {1995})}\BibitemShut {NoStop}%
\bibitem [{\citenamefont {Cirac}\ and\ \citenamefont {Zoller}(1995)}]{Cirac}%
  \BibitemOpen
  \bibfield  {author} {\bibinfo {author} {\bibfnamefont {J.~I.}\ \bibnamefont
  {Cirac}}\ and\ \bibinfo {author} {\bibfnamefont {P.}~\bibnamefont {Zoller}},\
  }\bibfield  {title} {\bibinfo {title} {Quantum computations with cold trapped
  ions},\ }\href@noop {} {\bibfield  {journal} {\bibinfo  {journal} {Phys. Rev.
  Lett.}\ }\textbf {\bibinfo {volume} {74}},\ \bibinfo {pages} {4091} (\bibinfo
  {year} {1995})}\BibitemShut {NoStop}%
\bibitem [{\citenamefont {Perdriat}\ \emph {et~al.}(2021)\citenamefont
  {Perdriat}, \citenamefont {Pellet-Mary}, \citenamefont {Huillery},
  \citenamefont {Rondin},\ and\ \citenamefont {Hétet}}]{Perdriat}%
  \BibitemOpen
  \bibfield  {author} {\bibinfo {author} {\bibfnamefont {M.}~\bibnamefont
  {Perdriat}}, \bibinfo {author} {\bibfnamefont {C.}~\bibnamefont
  {Pellet-Mary}}, \bibinfo {author} {\bibfnamefont {P.}~\bibnamefont
  {Huillery}}, \bibinfo {author} {\bibfnamefont {L.}~\bibnamefont {Rondin}},\
  and\ \bibinfo {author} {\bibfnamefont {G.}~\bibnamefont {Hétet}},\
  }\bibfield  {title} {\bibinfo {title} {Spin-mechanics with nitrogen-vacancy
  centers and trapped particles},\ }\bibfield  {journal} {\bibinfo  {journal}
  {Micromachines}\ }\textbf {\bibinfo {volume} {12}},\ \href
  {https://doi.org/10.3390/mi12060651} {10.3390/mi12060651} (\bibinfo {year}
  {2021})\BibitemShut {NoStop}%
\bibitem [{\citenamefont {Gonzalez-Ballestero}\ \emph
  {et~al.}(2021)\citenamefont {Gonzalez-Ballestero}, \citenamefont
  {Aspelmeyer}, \citenamefont {Novotny}, \citenamefont {Quidant},\ and\
  \citenamefont {Romero-Isart}}]{Gonzalez}%
  \BibitemOpen
  \bibfield  {author} {\bibinfo {author} {\bibfnamefont {C.}~\bibnamefont
  {Gonzalez-Ballestero}}, \bibinfo {author} {\bibfnamefont {M.}~\bibnamefont
  {Aspelmeyer}}, \bibinfo {author} {\bibfnamefont {L.}~\bibnamefont {Novotny}},
  \bibinfo {author} {\bibfnamefont {R.}~\bibnamefont {Quidant}},\ and\ \bibinfo
  {author} {\bibfnamefont {O.}~\bibnamefont {Romero-Isart}},\ }\bibfield
  {title} {\bibinfo {title} {Levitodynamics: Levitation and control of
  microscopic objects in vacuum},\ }\href
  {https://doi.org/10.1126/science.abg3027} {\bibfield  {journal} {\bibinfo
  {journal} {Science}\ }\textbf {\bibinfo {volume} {374}},\ \bibinfo {pages}
  {eabg3027} (\bibinfo {year} {2021})},\ \Eprint
  {https://arxiv.org/abs/https://www.science.org/doi/pdf/10.1126/science.abg3027}
  {https://www.science.org/doi/pdf/10.1126/science.abg3027} \BibitemShut
  {NoStop}%
\bibitem [{\citenamefont {Jackson~Kimball}\ \emph {et~al.}(2016)\citenamefont
  {Jackson~Kimball}, \citenamefont {Sushkov},\ and\ \citenamefont
  {Budker}}]{Kimball}%
  \BibitemOpen
  \bibfield  {author} {\bibinfo {author} {\bibfnamefont {D.~F.}\ \bibnamefont
  {Jackson~Kimball}}, \bibinfo {author} {\bibfnamefont {A.~O.}\ \bibnamefont
  {Sushkov}},\ and\ \bibinfo {author} {\bibfnamefont {D.}~\bibnamefont
  {Budker}},\ }\bibfield  {title} {\bibinfo {title} {Precessing ferromagnetic
  needle magnetometer},\ }\href
  {https://doi.org/10.1103/PhysRevLett.116.190801} {\bibfield  {journal}
  {\bibinfo  {journal} {Phys. Rev. Lett.}\ }\textbf {\bibinfo {volume} {116}},\
  \bibinfo {pages} {190801} (\bibinfo {year} {2016})}\BibitemShut {NoStop}%
\bibitem [{\citenamefont {Rusconi}\ \emph {et~al.}(2017)\citenamefont
  {Rusconi}, \citenamefont {P\"ochhacker}, \citenamefont {Kustura},
  \citenamefont {Cirac},\ and\ \citenamefont {Romero-Isart}}]{Rusconi}%
  \BibitemOpen
  \bibfield  {author} {\bibinfo {author} {\bibfnamefont {C.~C.}\ \bibnamefont
  {Rusconi}}, \bibinfo {author} {\bibfnamefont {V.}~\bibnamefont
  {P\"ochhacker}}, \bibinfo {author} {\bibfnamefont {K.}~\bibnamefont
  {Kustura}}, \bibinfo {author} {\bibfnamefont {J.~I.}\ \bibnamefont {Cirac}},\
  and\ \bibinfo {author} {\bibfnamefont {O.}~\bibnamefont {Romero-Isart}},\
  }\bibfield  {title} {\bibinfo {title} {Quantum spin stabilized magnetic
  levitation},\ }\href {https://doi.org/10.1103/PhysRevLett.119.167202}
  {\bibfield  {journal} {\bibinfo  {journal} {Phys. Rev. Lett.}\ }\textbf
  {\bibinfo {volume} {119}},\ \bibinfo {pages} {167202} (\bibinfo {year}
  {2017})}\BibitemShut {NoStop}%
\bibitem [{\citenamefont {Kustura}\ \emph {et~al.}(2022)\citenamefont
  {Kustura}, \citenamefont {Wachter}, \citenamefont {Rubio~L\'opez},\ and\
  \citenamefont {Rusconi}}]{Kustura}%
  \BibitemOpen
  \bibfield  {author} {\bibinfo {author} {\bibfnamefont {K.}~\bibnamefont
  {Kustura}}, \bibinfo {author} {\bibfnamefont {V.}~\bibnamefont {Wachter}},
  \bibinfo {author} {\bibfnamefont {A.~E.}\ \bibnamefont {Rubio~L\'opez}},\
  and\ \bibinfo {author} {\bibfnamefont {C.~C.}\ \bibnamefont {Rusconi}},\
  }\bibfield  {title} {\bibinfo {title} {Stability of a magnetically levitated
  nanomagnet in vacuum: Effects of gas and magnetization damping},\ }\href@noop
  {} {\bibfield  {journal} {\bibinfo  {journal} {Phys. Rev. B}\ }\textbf
  {\bibinfo {volume} {105}},\ \bibinfo {pages} {174439} (\bibinfo {year}
  {2022})}\BibitemShut {NoStop}%
\bibitem [{\citenamefont {Vinante}\ \emph {et~al.}(2021)\citenamefont
  {Vinante}, \citenamefont {Timberlake}, \citenamefont {Budker}, \citenamefont
  {Kimball}, \citenamefont {Sushkov},\ and\ \citenamefont
  {Ulbricht}}]{Vinante}%
  \BibitemOpen
  \bibfield  {author} {\bibinfo {author} {\bibfnamefont {A.}~\bibnamefont
  {Vinante}}, \bibinfo {author} {\bibfnamefont {C.}~\bibnamefont {Timberlake}},
  \bibinfo {author} {\bibfnamefont {D.}~\bibnamefont {Budker}}, \bibinfo
  {author} {\bibfnamefont {D.~F.~J.}\ \bibnamefont {Kimball}}, \bibinfo
  {author} {\bibfnamefont {A.~O.}\ \bibnamefont {Sushkov}},\ and\ \bibinfo
  {author} {\bibfnamefont {H.}~\bibnamefont {Ulbricht}},\ }\bibfield  {title}
  {\bibinfo {title} {Surpassing the energy resolution limit with ferromagnetic
  torque sensors},\ }\href@noop {} {\bibfield  {journal} {\bibinfo  {journal}
  {Phys. Rev. Lett.}\ }\textbf {\bibinfo {volume} {127}},\ \bibinfo {pages}
  {070801} (\bibinfo {year} {2021})}\BibitemShut {NoStop}%
\bibitem [{\citenamefont {Huillery}\ \emph {et~al.}(2020)\citenamefont
  {Huillery}, \citenamefont {Delord}, \citenamefont {Nicolas}, \citenamefont
  {Van Den~Bossche}, \citenamefont {Perdriat},\ and\ \citenamefont
  {H\'etet}}]{Huillery}%
  \BibitemOpen
  \bibfield  {author} {\bibinfo {author} {\bibfnamefont {P.}~\bibnamefont
  {Huillery}}, \bibinfo {author} {\bibfnamefont {T.}~\bibnamefont {Delord}},
  \bibinfo {author} {\bibfnamefont {L.}~\bibnamefont {Nicolas}}, \bibinfo
  {author} {\bibfnamefont {M.}~\bibnamefont {Van Den~Bossche}}, \bibinfo
  {author} {\bibfnamefont {M.}~\bibnamefont {Perdriat}},\ and\ \bibinfo
  {author} {\bibfnamefont {G.}~\bibnamefont {H\'etet}},\ }\bibfield  {title}
  {\bibinfo {title} {Spin mechanics with levitating ferromagnetic particles},\
  }\href {https://doi.org/10.1103/PhysRevB.101.134415} {\bibfield  {journal}
  {\bibinfo  {journal} {Phys. Rev. B}\ }\textbf {\bibinfo {volume} {101}},\
  \bibinfo {pages} {134415} (\bibinfo {year} {2020})}\BibitemShut {NoStop}%
\bibitem [{\citenamefont {Gieseler}\ \emph {et~al.}(2020)\citenamefont
  {Gieseler}, \citenamefont {Kabcenell}, \citenamefont {Rosenfeld},
  \citenamefont {Schaefer}, \citenamefont {Safira}, \citenamefont {Schuetz},
  \citenamefont {Gonzalez-Ballestero}, \citenamefont {Rusconi}, \citenamefont
  {Romero-Isart},\ and\ \citenamefont {Lukin}}]{Gieseler3}%
  \BibitemOpen
  \bibfield  {author} {\bibinfo {author} {\bibfnamefont {J.}~\bibnamefont
  {Gieseler}}, \bibinfo {author} {\bibfnamefont {A.}~\bibnamefont {Kabcenell}},
  \bibinfo {author} {\bibfnamefont {E.}~\bibnamefont {Rosenfeld}}, \bibinfo
  {author} {\bibfnamefont {J.~D.}\ \bibnamefont {Schaefer}}, \bibinfo {author}
  {\bibfnamefont {A.}~\bibnamefont {Safira}}, \bibinfo {author} {\bibfnamefont
  {M.~J.~A.}\ \bibnamefont {Schuetz}}, \bibinfo {author} {\bibfnamefont
  {C.}~\bibnamefont {Gonzalez-Ballestero}}, \bibinfo {author} {\bibfnamefont
  {C.~C.}\ \bibnamefont {Rusconi}}, \bibinfo {author} {\bibfnamefont
  {O.}~\bibnamefont {Romero-Isart}},\ and\ \bibinfo {author} {\bibfnamefont
  {M.~D.}\ \bibnamefont {Lukin}},\ }\bibfield  {title} {\bibinfo {title}
  {Single-spin magnetomechanics with levitated micromagnets},\ }\href
  {https://doi.org/10.1103/PhysRevLett.124.163604} {\bibfield  {journal}
  {\bibinfo  {journal} {Phys. Rev. Lett.}\ }\textbf {\bibinfo {volume} {124}},\
  \bibinfo {pages} {163604} (\bibinfo {year} {2020})}\BibitemShut {NoStop}%
\bibitem [{\citenamefont {Timberlake}\ \emph {et~al.}(2021)\citenamefont
  {Timberlake}, \citenamefont {Vinante}, \citenamefont {Shankar}, \citenamefont
  {Lapi},\ and\ \citenamefont {Ulbricht}}]{Timberlake}%
  \BibitemOpen
  \bibfield  {author} {\bibinfo {author} {\bibfnamefont {C.}~\bibnamefont
  {Timberlake}}, \bibinfo {author} {\bibfnamefont {A.}~\bibnamefont {Vinante}},
  \bibinfo {author} {\bibfnamefont {F.}~\bibnamefont {Shankar}}, \bibinfo
  {author} {\bibfnamefont {A.}~\bibnamefont {Lapi}},\ and\ \bibinfo {author}
  {\bibfnamefont {H.}~\bibnamefont {Ulbricht}},\ }\bibfield  {title} {\bibinfo
  {title} {Probing modified gravity with magnetically levitated resonators},\
  }\href@noop {} {\bibfield  {journal} {\bibinfo  {journal} {Phys. Rev. D}\
  }\textbf {\bibinfo {volume} {104}},\ \bibinfo {pages} {L101101} (\bibinfo
  {year} {2021})}\BibitemShut {NoStop}%
\bibitem [{\citenamefont {Simon}\ \emph {et~al.}(1997)\citenamefont {Simon},
  \citenamefont {Heflinger},\ and\ \citenamefont {Ridgway}}]{levitron}%
  \BibitemOpen
  \bibfield  {author} {\bibinfo {author} {\bibfnamefont {M.~D.}\ \bibnamefont
  {Simon}}, \bibinfo {author} {\bibfnamefont {L.~O.}\ \bibnamefont
  {Heflinger}},\ and\ \bibinfo {author} {\bibfnamefont {S.~L.}\ \bibnamefont
  {Ridgway}},\ }\bibfield  {title} {\bibinfo {title} {Spin stabilized magnetic
  levitation},\ }\href@noop {} {\bibfield  {journal} {\bibinfo  {journal}
  {American Journal of Physics}\ }\textbf {\bibinfo {volume} {65}},\ \bibinfo
  {pages} {286} (\bibinfo {year} {1997})}\BibitemShut {NoStop}%
\bibitem [{\citenamefont {Bachelet}(1991)}]{maglev}%
  \BibitemOpen
  \bibfield  {author} {\bibinfo {author} {\bibfnamefont {E.}~\bibnamefont
  {Bachelet}},\ }\href@noop {} {\bibfield  {journal} {\bibinfo  {journal}
  {Levitating transmitting apparatus, US patent 1020943A, 6th of Dec.}\ }
  (\bibinfo {year} {1991})}\BibitemShut {NoStop}%
\bibitem [{\citenamefont {Druge}\ \emph {et~al.}(2014)\citenamefont {Druge},
  \citenamefont {Jean}, \citenamefont {Laurent}, \citenamefont
  {M{\'{e}}asson},\ and\ \citenamefont {Favero}}]{Druge_2014}%
  \BibitemOpen
  \bibfield  {author} {\bibinfo {author} {\bibfnamefont {J.}~\bibnamefont
  {Druge}}, \bibinfo {author} {\bibfnamefont {C.}~\bibnamefont {Jean}},
  \bibinfo {author} {\bibfnamefont {O.}~\bibnamefont {Laurent}}, \bibinfo
  {author} {\bibfnamefont {M.-A.}\ \bibnamefont {M{\'{e}}asson}},\ and\
  \bibinfo {author} {\bibfnamefont {I.}~\bibnamefont {Favero}},\ }\bibfield
  {title} {\bibinfo {title} {Damping and non-linearity of a levitating magnet
  in rotation above a superconductor},\ }\href
  {https://doi.org/10.1088/1367-2630/16/7/075011} {\bibfield  {journal}
  {\bibinfo  {journal} {New Journal of Physics}\ }\textbf {\bibinfo {volume}
  {16}},\ \bibinfo {pages} {075011} (\bibinfo {year} {2014})}\BibitemShut
  {NoStop}%
\bibitem [{\citenamefont {Nemoshkalenko}\ \emph {et~al.}(1990)\citenamefont
  {Nemoshkalenko}, \citenamefont {Brandt}, \citenamefont {Kordyuk},\ and\
  \citenamefont {Nikitin}}]{NEMOSHKALENKO1990481}%
  \BibitemOpen
  \bibfield  {author} {\bibinfo {author} {\bibfnamefont {V.}~\bibnamefont
  {Nemoshkalenko}}, \bibinfo {author} {\bibfnamefont {E.}~\bibnamefont
  {Brandt}}, \bibinfo {author} {\bibfnamefont {A.}~\bibnamefont {Kordyuk}},\
  and\ \bibinfo {author} {\bibfnamefont {B.}~\bibnamefont {Nikitin}},\
  }\bibfield  {title} {\bibinfo {title} {Dynamics of a permanent magnet
  levitating above a high-tc superconductor},\ }\href
  {https://doi.org/https://doi.org/10.1016/0921-4534(90)90019-B} {\bibfield
  {journal} {\bibinfo  {journal} {Physica C: Superconductivity}\ }\textbf
  {\bibinfo {volume} {170}},\ \bibinfo {pages} {481} (\bibinfo {year}
  {1990})}\BibitemShut {NoStop}%
\bibitem [{\citenamefont {Hull}\ and\ \citenamefont {Cansiz}(1999)}]{Hull}%
  \BibitemOpen
  \bibfield  {author} {\bibinfo {author} {\bibfnamefont {J.~R.}\ \bibnamefont
  {Hull}}\ and\ \bibinfo {author} {\bibfnamefont {A.}~\bibnamefont {Cansiz}},\
  }\bibfield  {title} {\bibinfo {title} {Vertical and lateral forces between a
  permanent magnet and a high-temperature superconductor},\ }\href@noop {}
  {\bibfield  {journal} {\bibinfo  {journal} {Journal of Applied Physics}\
  }\textbf {\bibinfo {volume} {86}},\ \bibinfo {pages} {6396} (\bibinfo {year}
  {1999})}\BibitemShut {NoStop}%
\bibitem [{\citenamefont {Vinante}\ \emph {et~al.}(2020)\citenamefont
  {Vinante}, \citenamefont {Falferi}, \citenamefont {Gasbarri}, \citenamefont
  {Setter}, \citenamefont {Timberlake},\ and\ \citenamefont
  {Ulbricht}}]{Vinante3}%
  \BibitemOpen
  \bibfield  {author} {\bibinfo {author} {\bibfnamefont {A.}~\bibnamefont
  {Vinante}}, \bibinfo {author} {\bibfnamefont {P.}~\bibnamefont {Falferi}},
  \bibinfo {author} {\bibfnamefont {G.}~\bibnamefont {Gasbarri}}, \bibinfo
  {author} {\bibfnamefont {A.}~\bibnamefont {Setter}}, \bibinfo {author}
  {\bibfnamefont {C.}~\bibnamefont {Timberlake}},\ and\ \bibinfo {author}
  {\bibfnamefont {H.}~\bibnamefont {Ulbricht}},\ }\bibfield  {title} {\bibinfo
  {title} {Ultralow mechanical damping with meissner-levitated ferromagnetic
  microparticles},\ }\href {https://doi.org/10.1103/PhysRevApplied.13.064027}
  {\bibfield  {journal} {\bibinfo  {journal} {Phys. Rev. Applied}\ }\textbf
  {\bibinfo {volume} {13}},\ \bibinfo {pages} {064027} (\bibinfo {year}
  {2020})}\BibitemShut {NoStop}%
\bibitem [{\citenamefont {Sackett}\ \emph {et~al.}(1993)\citenamefont
  {Sackett}, \citenamefont {Cornell}, \citenamefont {Monroe},\ and\
  \citenamefont {Wieman}}]{Sackett}%
  \BibitemOpen
  \bibfield  {author} {\bibinfo {author} {\bibfnamefont {C.}~\bibnamefont
  {Sackett}}, \bibinfo {author} {\bibfnamefont {E.}~\bibnamefont {Cornell}},
  \bibinfo {author} {\bibfnamefont {C.}~\bibnamefont {Monroe}},\ and\ \bibinfo
  {author} {\bibfnamefont {C.}~\bibnamefont {Wieman}},\ }\bibfield  {title}
  {\bibinfo {title} {A magnetic suspension system for atoms and bar magnets},\
  }\href@noop {} {\bibfield  {journal} {\bibinfo  {journal} {American Journal
  of Physics}\ }\textbf {\bibinfo {volume} {61}},\ \bibinfo {pages} {304}
  (\bibinfo {year} {1993})}\BibitemShut {NoStop}%
\bibitem [{\citenamefont {Cornell}\ \emph {et~al.}(1991)\citenamefont
  {Cornell}, \citenamefont {Monroe},\ and\ \citenamefont
  {Wieman}}]{CornelNeutral}%
  \BibitemOpen
  \bibfield  {author} {\bibinfo {author} {\bibfnamefont {E.~A.}\ \bibnamefont
  {Cornell}}, \bibinfo {author} {\bibfnamefont {C.}~\bibnamefont {Monroe}},\
  and\ \bibinfo {author} {\bibfnamefont {C.~E.}\ \bibnamefont {Wieman}},\
  }\bibfield  {title} {\bibinfo {title} {Multiply loaded, ac magnetic trap for
  neutral atoms},\ }\href {https://doi.org/10.1103/PhysRevLett.67.2439}
  {\bibfield  {journal} {\bibinfo  {journal} {Phys. Rev. Lett.}\ }\textbf
  {\bibinfo {volume} {67}},\ \bibinfo {pages} {2439} (\bibinfo {year}
  {1991})}\BibitemShut {NoStop}%
\bibitem [{\citenamefont {BASSANI}(2005)}]{BASSANI}%
  \BibitemOpen
  \bibfield  {author} {\bibinfo {author} {\bibfnamefont {R.}~\bibnamefont
  {BASSANI}},\ }\bibfield  {title} {\bibinfo {title} {Stability of permanent
  magnet bearings under parametric excitations},\ }\href@noop {} {\bibfield
  {journal} {\bibinfo  {journal} {Tribology Transactions}\ }\textbf {\bibinfo
  {volume} {48}},\ \bibinfo {pages} {457} (\bibinfo {year} {2005})}\BibitemShut
  {NoStop}%
\bibitem [{\citenamefont {Bassani}(2006)}]{BASSANI2}%
  \BibitemOpen
  \bibfield  {author} {\bibinfo {author} {\bibfnamefont {R.}~\bibnamefont
  {Bassani}},\ }\bibfield  {title} {\bibinfo {title} {{A Stability Space of a
  Magnetomechanical Bearing}},\ }\href {https://doi.org/10.1115/1.2431812}
  {\bibfield  {journal} {\bibinfo  {journal} {Journal of Dynamic Systems,
  Measurement, and Control}\ }\textbf {\bibinfo {volume} {129}},\ \bibinfo
  {pages} {178} (\bibinfo {year} {2006})},\ \Eprint
  {https://arxiv.org/abs/https://asmedigitalcollection.asme.org/dynamicsystems/article-pdf/129/2/178/5526944/178\_1.pdf}
  {https://asmedigitalcollection.asme.org/dynamicsystems/article-pdf/129/2/178/5526944/178\_1.pdf}
  \BibitemShut {NoStop}%
\bibitem [{\citenamefont {Bassani}(2007)}]{BASSANI3}%
  \BibitemOpen
  \bibfield  {author} {\bibinfo {author} {\bibfnamefont {R.}~\bibnamefont
  {Bassani}},\ }\bibfield  {title} {\bibinfo {title} {Dynamic stability of
  passive magnetic bearings},\ }\href@noop {} {\bibfield  {journal} {\bibinfo
  {journal} {Nonlinear Dynamics}\ }\textbf {\bibinfo {volume} {50}},\ \bibinfo
  {pages} {161} (\bibinfo {year} {2007})}\BibitemShut {NoStop}%
\bibitem [{\citenamefont {Folman}\ \emph {et~al.}(2000)\citenamefont {Folman},
  \citenamefont {Kr\"uger}, \citenamefont {Cassettari}, \citenamefont {Hessmo},
  \citenamefont {Maier},\ and\ \citenamefont
  {Schmiedmayer}}]{PhysRevLett.84.4749}%
  \BibitemOpen
  \bibfield  {author} {\bibinfo {author} {\bibfnamefont {R.}~\bibnamefont
  {Folman}}, \bibinfo {author} {\bibfnamefont {P.}~\bibnamefont {Kr\"uger}},
  \bibinfo {author} {\bibfnamefont {D.}~\bibnamefont {Cassettari}}, \bibinfo
  {author} {\bibfnamefont {B.}~\bibnamefont {Hessmo}}, \bibinfo {author}
  {\bibfnamefont {T.}~\bibnamefont {Maier}},\ and\ \bibinfo {author}
  {\bibfnamefont {J.}~\bibnamefont {Schmiedmayer}},\ }\bibfield  {title}
  {\bibinfo {title} {Controlling cold atoms using nanofabricated surfaces: Atom
  chips},\ }\href {https://doi.org/10.1103/PhysRevLett.84.4749} {\bibfield
  {journal} {\bibinfo  {journal} {Phys. Rev. Lett.}\ }\textbf {\bibinfo
  {volume} {84}},\ \bibinfo {pages} {4749} (\bibinfo {year}
  {2000})}\BibitemShut {NoStop}%
\bibitem [{\citenamefont {Feenstra}\ \emph {et~al.}(2004)\citenamefont
  {Feenstra}, \citenamefont {Andersson},\ and\ \citenamefont
  {Schmiedmayer}}]{Feenstra}%
  \BibitemOpen
  \bibfield  {author} {\bibinfo {author} {\bibfnamefont {L.}~\bibnamefont
  {Feenstra}}, \bibinfo {author} {\bibfnamefont {L.~M.}\ \bibnamefont
  {Andersson}},\ and\ \bibinfo {author} {\bibfnamefont {J.}~\bibnamefont
  {Schmiedmayer}},\ }\bibfield  {title} {\bibinfo {title} {Microtraps and atom
  chips: Toolboxes for cold atom physics},\ }\href@noop {} {\bibfield
  {journal} {\bibinfo  {journal} {General Relativity and Gravitation}\ }\textbf
  {\bibinfo {volume} {36}},\ \bibinfo {pages} {2317} (\bibinfo {year}
  {2004})}\BibitemShut {NoStop}%
\bibitem [{\citenamefont {Rahman}\ \emph {et~al.}(2016)\citenamefont {Rahman},
  \citenamefont {Frangeskou}, \citenamefont {Kim}, \citenamefont {Bose},
  \citenamefont {Morley},\ and\ \citenamefont {Barker}}]{Rahman}%
  \BibitemOpen
  \bibfield  {author} {\bibinfo {author} {\bibfnamefont {A.~T. M.~A.}\
  \bibnamefont {Rahman}}, \bibinfo {author} {\bibfnamefont {A.~C.}\
  \bibnamefont {Frangeskou}}, \bibinfo {author} {\bibfnamefont {M.~S.}\
  \bibnamefont {Kim}}, \bibinfo {author} {\bibfnamefont {S.}~\bibnamefont
  {Bose}}, \bibinfo {author} {\bibfnamefont {G.~W.}\ \bibnamefont {Morley}},\
  and\ \bibinfo {author} {\bibfnamefont {P.~F.}\ \bibnamefont {Barker}},\
  }\bibfield  {title} {\bibinfo {title} {Burning and graphitization of
  optically levitated nanodiamonds in vacuum},\ }\href@noop {} {\bibfield
  {journal} {\bibinfo  {journal} {Scientific Reports}\ }\textbf {\bibinfo
  {volume} {6}},\ \bibinfo {pages} {21633 EP } (\bibinfo {year}
  {2016})}\BibitemShut {NoStop}%
\bibitem [{\citenamefont {Neukirch}\ \emph {et~al.}(2013)\citenamefont
  {Neukirch}, \citenamefont {Gieseler}, \citenamefont {Quidant}, \citenamefont
  {Novotny},\ and\ \citenamefont {Nick~Vamivakas}}]{Neukirch}%
  \BibitemOpen
  \bibfield  {author} {\bibinfo {author} {\bibfnamefont {L.~P.}\ \bibnamefont
  {Neukirch}}, \bibinfo {author} {\bibfnamefont {J.}~\bibnamefont {Gieseler}},
  \bibinfo {author} {\bibfnamefont {R.}~\bibnamefont {Quidant}}, \bibinfo
  {author} {\bibfnamefont {L.}~\bibnamefont {Novotny}},\ and\ \bibinfo {author}
  {\bibfnamefont {A.}~\bibnamefont {Nick~Vamivakas}},\ }\bibfield  {title}
  {\bibinfo {title} {Observation of nitrogen vacancy photoluminescence from an
  optically levitated nanodiamond},\ }\href@noop {} {\bibfield  {journal}
  {\bibinfo  {journal} {Optics Letters}\ }\textbf {\bibinfo {volume} {38}},\
  \bibinfo {pages} {2976} (\bibinfo {year} {2013})}\BibitemShut {NoStop}%
\bibitem [{\citenamefont {Hoang}\ \emph {et~al.}(2016)\citenamefont {Hoang},
  \citenamefont {Ahn}, \citenamefont {Bang},\ and\ \citenamefont {Li}}]{Hoang}%
  \BibitemOpen
  \bibfield  {author} {\bibinfo {author} {\bibfnamefont {T.~M.}\ \bibnamefont
  {Hoang}}, \bibinfo {author} {\bibfnamefont {J.}~\bibnamefont {Ahn}}, \bibinfo
  {author} {\bibfnamefont {J.}~\bibnamefont {Bang}},\ and\ \bibinfo {author}
  {\bibfnamefont {T.}~\bibnamefont {Li}},\ }\bibfield  {title} {\bibinfo
  {title} {Electron spin control of optically levitated nanodiamonds in
  vacuum},\ }\href@noop {} {\bibfield  {journal} {\bibinfo  {journal} {Nature
  Communications}\ }\textbf {\bibinfo {volume} {7}},\ \bibinfo {pages} {12250
  EP } (\bibinfo {year} {2016})}\BibitemShut {NoStop}%
\end{thebibliography}%


\begin{thebibliography}{2}%
\makeatletter
\providecommand \@ifxundefined [1]{%
 \@ifx{#1\undefined}
}%
\providecommand \@ifnum [1]{%
 \ifnum #1\expandafter \@firstoftwo
 \else \expandafter \@secondoftwo
 \fi
}%
\providecommand \@ifx [1]{%
 \ifx #1\expandafter \@firstoftwo
 \else \expandafter \@secondoftwo
 \fi
}%
\providecommand \natexlab [1]{#1}%
\providecommand \enquote  [1]{``#1''}%
\providecommand \bibnamefont  [1]{#1}%
\providecommand \bibfnamefont [1]{#1}%
\providecommand \citenamefont [1]{#1}%
\providecommand \href@noop [0]{\@secondoftwo}%
\providecommand \href [0]{\begingroup \@sanitize@url \@href}%
\providecommand \@href[1]{\@@startlink{#1}\@@href}%
\providecommand \@@href[1]{\endgroup#1\@@endlink}%
\providecommand \@sanitize@url [0]{\catcode `\\12\catcode `\$12\catcode
  `\&12\catcode `\#12\catcode `\^12\catcode `\_12\catcode `\%12\relax}%
\providecommand \@@startlink[1]{}%
\providecommand \@@endlink[0]{}%
\providecommand \url  [0]{\begingroup\@sanitize@url \@url }%
\providecommand \@url [1]{\endgroup\@href {#1}{\urlprefix }}%
\providecommand \urlprefix  [0]{URL }%
\providecommand \Eprint [0]{\href }%
\providecommand \doibase [0]{https://doi.org/}%
\providecommand \selectlanguage [0]{\@gobble}%
\providecommand \bibinfo  [0]{\@secondoftwo}%
\providecommand \bibfield  [0]{\@secondoftwo}%
\providecommand \translation [1]{[#1]}%
\providecommand \BibitemOpen [0]{}%
\providecommand \bibitemStop [0]{}%
\providecommand \bibitemNoStop [0]{.\EOS\space}%
\providecommand \EOS [0]{\spacefactor3000\relax}%
\providecommand \BibitemShut  [1]{\csname bibitem#1\endcsname}%
\let\auto@bib@innerbib\@empty
\bibitem [{\citenamefont {Kirillov}\ and\ \citenamefont
  {Levi}(2017)}]{Kirillov}%
  \BibitemOpen
  \bibfield  {author} {\bibinfo {author} {\bibfnamefont {O.}~\bibnamefont
  {Kirillov}}\ and\ \bibinfo {author} {\bibfnamefont {M.}~\bibnamefont
  {Levi}},\ }\bibfield  {title} {\bibinfo {title} {A coriolis force in an
  inertial frame},\ }\href@noop {} {\ \textbf {\bibinfo {volume} {30}},\
  \bibinfo {pages} {1109} (\bibinfo {year} {2017})}\BibitemShut {NoStop}%
\bibitem [{\citenamefont {Dehmelt}\ \emph {et~al.}(1968)\citenamefont
  {Dehmelt}, \citenamefont {Bates},\ and\ \citenamefont {Estermann}}]{Dehmelt}%
  \BibitemOpen
  \bibfield  {author} {\bibinfo {author} {\bibfnamefont {H.~G.}\ \bibnamefont
  {Dehmelt}}, \bibinfo {author} {\bibfnamefont {D.~R.}\ \bibnamefont {Bates}},\
  and\ \bibinfo {author} {\bibfnamefont {I.}~\bibnamefont {Estermann}},\
  }\bibinfo {title} {Radiofrequency spectroscopy of stored ions i:
  Storage**part ii: Spectroscopy is now scheduled to appear in volume v of this
  series.}\ (\bibinfo  {publisher} {Academic Press},\ \bibinfo {year} {1968})\
  pp.\ \bibinfo {pages} {53--72}\BibitemShut {NoStop}%
\end{thebibliography}%

\end{document}


\vspace{0.2in}
{\Large \hspace{1.6in}\textsc{Supplemental Material} }\\

\title{Planar Magnetic Paul Traps for Ferromagnetic Particles}


\author{M. Perdriat, C. Pellet-Mary, T. Copie, G. H\'etet$^{1}$} 
\affiliation{$^1$Laboratoire De Physique de l'\'Ecole Normale Sup\'erieure, \'Ecole Normale Sup\'erieure, PSL Research University, CNRS, Sorbonne Universit\'e, Universit\'e de Paris Cit\'e, 24 rue Lhomond, 75231 Paris Cedex 05, France}
\maketitle

\tableofcontents

\section{Rotating magnetic saddle}

In this section, we explain the origin of the confinement of a particle subjected to a rotating saddle potential. We derive the stability condition and the macromotion equation as well as the resulting secular frequency. Last, we include particle size effects in order to explain the $z$-confinement.

\subsection{The rotating saddle}

We first consider a particle of mass $m$ which can move in a plane subjected to a saddle potential (non-magnetic at first) rotating at the angular frequency $\omega$. We designate by $(x,y)$ the spatial coordinate of the particle in the laboratory-fixed frame and $(X,Y)$ the coordinate in the rotating saddle frame.  The rotating saddle potential reads:
\begin{align}
U(X,Y)=\frac{1}{2}m\omega_r^2\left(X^2-Y^2\right).
\end{align}
Writing the motion equation of the particle in the rotating frame, one adds the centrifugal force $\mathbf{F}_{\rm cen}=m\Omega^2 (X\mathbf{e}_X+Y \mathbf{e}_Y)$ and the Coriolis force $\mathbf{F}_{\rm Cor}=-2m(\Omega \mathbf{e}_Z) \times (\dot{X}\mathbf{e}_X+\dot{Y} \mathbf{e}_Y) $.
The fundamental principle of dynamics gives the coupled set of equations:
\begin{align}
\ddot{X}-2\Omega\dot{Y}+\left(\omega_r^2-\Omega^2 \right)X&=0, \\
\ddot{Y}+2\Omega\dot{X}-\left(\omega_r^2+\Omega^2 \right)Y&=0.
\end{align}
This set of equations can be written as $\mathbf{\dot{U}}=\mathbf{A}\mathbf{U}$ with $\mathbf{U}= \,^t(X,Y,\dot{X},\dot{Y})$ and
\begin{align}
\mathbf{A}=\begin{pmatrix}
0&0&1&0\\
0&0&0&1\\
\Omega^2-\omega_r^2&0&0&2\Omega\\
0&\omega_r^2+\Omega^2&-2\Omega&0
\end{pmatrix}.
\end{align}
The particle is stable if and only if the eigenvalues of the matrix $\mathbf{A}$ have a negative real part. The eigenvalues verify the equality:
\begin{align}
\lambda^4+2\Omega^2\lambda^2+\Omega^4-\omega_r^4=0,
\end{align}
and therefore:
\begin{align}
\lambda^2= \pm \omega_r^2-\Omega^2.
\end{align}
The eigenvalues have no positive real part if and only if $\Omega-\omega_r>0$. Thus, we obtain the stability condition:
\begin{align}
\Omega>\omega_r.
\end{align}
To estimate the secular frequency, we will employ the approach proposed in \cite{Kirillov}. The motional equation of the particle in the laboratory-fixed frame reads:
\begin{align}
\mathbf{\ddot{V}}+\omega_r^2 \mathbf{S}(\Omega t)\mathbf{V}=0,
\end{align}
with
\begin{align}
\mathbf{V}&=\,^t(x,y),\\
\mathbf{S}(\Omega t)&=\begin{pmatrix}
\cos{(2\Omega t)} & \sin{(2\Omega t)}\\
\sin{(2\Omega t)} & -\cos{(2\Omega t)}
\end{pmatrix}.
\end{align}

Using the transformation of the guiding-center \cite{Kirillov}:
\begin{align}
\mathbf{W}=\mathbf{V}-\frac{1}{4}\left( \frac{\omega_r}{\Omega}\right)^2 \mathbf{S}(\omega t) \left( \mathbf{V}-\frac{1}{\Omega}\mathbf{J}\mathbf{\dot{V}}\right), \mathbf{J}=\begin{pmatrix}
0&-1\\1&0
\end{pmatrix},
\end{align}
we obtain a differential equation for $\mathbf{W}$:
\begin{align}
\mathbf{\ddot{W}}-\frac{1}{4}\omega_r \left(\frac{\omega_r}{\Omega}\right)^3\mathbf{J}\mathbf{\dot{W}}+\frac{1}{4}\omega_r^2\left(\frac{\omega_r}{\Omega}\right)^2\mathbf{W}=\left(\frac{\omega_r}{\Omega} \right)^4 f\left(\omega_r^2\mathbf{W},\omega_r\mathbf{\dot{W}},\frac{\omega_r}{\Omega}\right),
\end{align}
where $f$ a linear function in $\omega_r^2\mathbf{W}$, $\omega_r\mathbf{\dot{W}}$ and analytic in $\omega_r/\Omega$ in a fixed neighborhood of $\omega_r/\Omega=0$.
In the limit $\omega_r/\Omega \rightarrow 0$, this equation results in a radial motion given by the characteristic confining secular frequency $\Tilde{\omega}=(2\pi)\frac{1}{2}\frac{\omega_r^2}{\Omega}$ and by a precessional motion at the characteristic secular frequency $\Tilde{\omega}_{\rm prec}=\frac{1}{4} \frac{\omega_r^4}{\Omega^3}$.

\subsection{The rotating magnetic saddle}

We now consider the rotating magnetic saddle described in the main text. We model the levitating magnet as a parallelepiped of length $l$, square cross-section with a side length $h$ and volume $V=l \times h^2 $. We suppose that the magnetic dipole of the magnet is oriented along $z$. The vector $\mathbf{R}=(X,Y,Z)$ designates the spatial coordinate in the saddle rotating frame and by $\mathbf{r}=(x,y,z)$ the spatial coordinate in the laboratory-fixed frame. The component along the $Z$ direction of the magnetic saddle in the rotating frame reads :
\begin{align}
\begin{split}
    B_{Z}\left(\mathbf{R} \right)&=\frac{B_Z''(Z)}{2}\left(X^2-Y^2\right),
\end{split}.
\end{align} 
to second order in $X,Y$.
Here $B_Z''$ is an even function of $Z$.
In the laboratory-fixed frame, this potential is time-dependent and reads:
\begin{align}
\begin{split}
    B_{z}(\mathbf{r},t)&=\frac{B_z''(z)}{2}\left((x^2-y^2)\cos{(2\Omega t)}-2xy\sin{(2\Omega t)} \right)
\end{split}.
\end{align} 
where $\Omega/2\pi$ is the rotation frequency of the magnetic saddle.

We suppose that an external homogeneous magnetic field  aligns the orientation of the magnet in the $z$ direction and that the angle along the $z$ axis is also confined. Experimentally, the confinement  is realized using an external permanent magnet. We can neglect the influence of the magnetic field components in the $x$ and $y$ directions since it is perpendicular to the dipole of the magnet. The magnetic energy of an infinitesimal element of volume $\rm dV$ then equals to ${\rm d}E_{\rm mag}(\mathbf{r},t)=-\mathbf{M} \cdot \mathbf{B}(\mathbf{r},t){\rm dV} =-M_zB_z(\mathbf{r},t){\rm dV}$. We designate by $(x,y,z)$ the spatial coordinate of the center of mass of the magnet. Integrating the magnetic energy over all the magnet volume, we obtain the total energy:
\begin{align}\label{totalenergy}
E_{\rm mag}(\mathbf{r},t)=\int_{-\frac{h}{2}+x}^{\frac{h}{2}+x}\int_{-\frac{l}{2}+y}^{\frac{l}{2}+y}\int_{-\frac{h}{2}+z}^{\frac{h}{2}+z} -\mathbf{M} \cdot \mathbf{B} {\rm dV},
\end{align}
which yields
\begin{align}
E_{\rm mag}(\mathbf{r},t)=-MV \left(\int_{-\frac{h}{2}+z}^{\frac{h}{2}+z}\frac{B_z''(z')}{2h}{\rm d}z' \right)\left(\cos{(2\Omega t)}\left(\frac{1}{3}\left(\frac{h^2}{4}-\frac{l^2}{4} \right)+x^2-y^2 \right) - 2\sin{(2\Omega t)}xy \right).
\end{align}
We define the function $F(z)=\int_{-\frac{h}{2}+z}^{\frac{h}{2}+z}\frac{B_z''(z')}{2h}{\rm d}z'$ which is a even function of $z$ because $B_z''$ is  a even function of $z$. 
We assume that the Taylor expansion of $F$ at second order in $z$ is verified in the parameter values of $z$ explored such that $F(z) \approx a_0 + a_2 \frac{z^2}{2}$. Keeping only the second order terms in the spatial coordinates, we obtain:
\begin{align}
E_{\rm mag}(\mathbf{r},t) \approx -MV a_0 \left(\cos{(2\Omega t)}\left(x^2-y^2 \right) - 2\sin{(2\Omega t)}xy \right) -MV \frac{a_2}{24}\left(h^2-l^2 \right)\cos{(2\Omega t)} z^2.
\end{align}
We obtain the energy:
\begin{align}
E_{\rm mag}(\mathbf{r},t) \approx \frac{1}{2}m \omega_r^2\left(\cos{(2\Omega t)}\left(y^2-x^2 \right) + 2\sin{(2\Omega t)}xy \right) +\frac{1}{2}m \omega_z^2 \cos{(2\Omega t)} z^2,
\end{align}
where we introduced the relevant frequencies:
\begin{align}
\omega_r/2\pi&=\sqrt{\frac{2B_{\rm sat}a_0}{\mu_0 \rho_m}},\\
\omega_z/2\pi&=\sqrt{\frac{B_{\rm sat}a_2}{12 \mu_0 \rho_m}\left(l^2-h^2\right)}.
\end{align}

In our experiment, the magnetic fields generated by the trap are strong enough so the magnets are trapped outside the harmonic region along $z$.
The position of the magnet in the $z$ direction is the result of a balance between the outward force from the rotating platform and gravity.
The pseudo- magnetic potential $\Psi_z$ along $z$, namely the magnetic energy averaged over one cycle of the trap rotation reads \cite{Dehmelt}:
\begin{eqnarray}
\Psi_z=\frac{| \nabla_z (\mu B_z(z)) |^2 }{4 m \Omega^2},
\end{eqnarray}
where $\mu=B_{\rm sat} V / \mu_0$ is the magnetic moment of the magnet. $B_z(z)$ is the time independent prefactor in the energy. Here, using \ref{totalenergy}, we find  
\begin{eqnarray}
B_z(z)=- M V F(z) \frac{1}{12}(h^2-l^2).
\end{eqnarray}

\begin{figure}[ht!]
\centerline{\scalebox{0.6}{\includegraphics{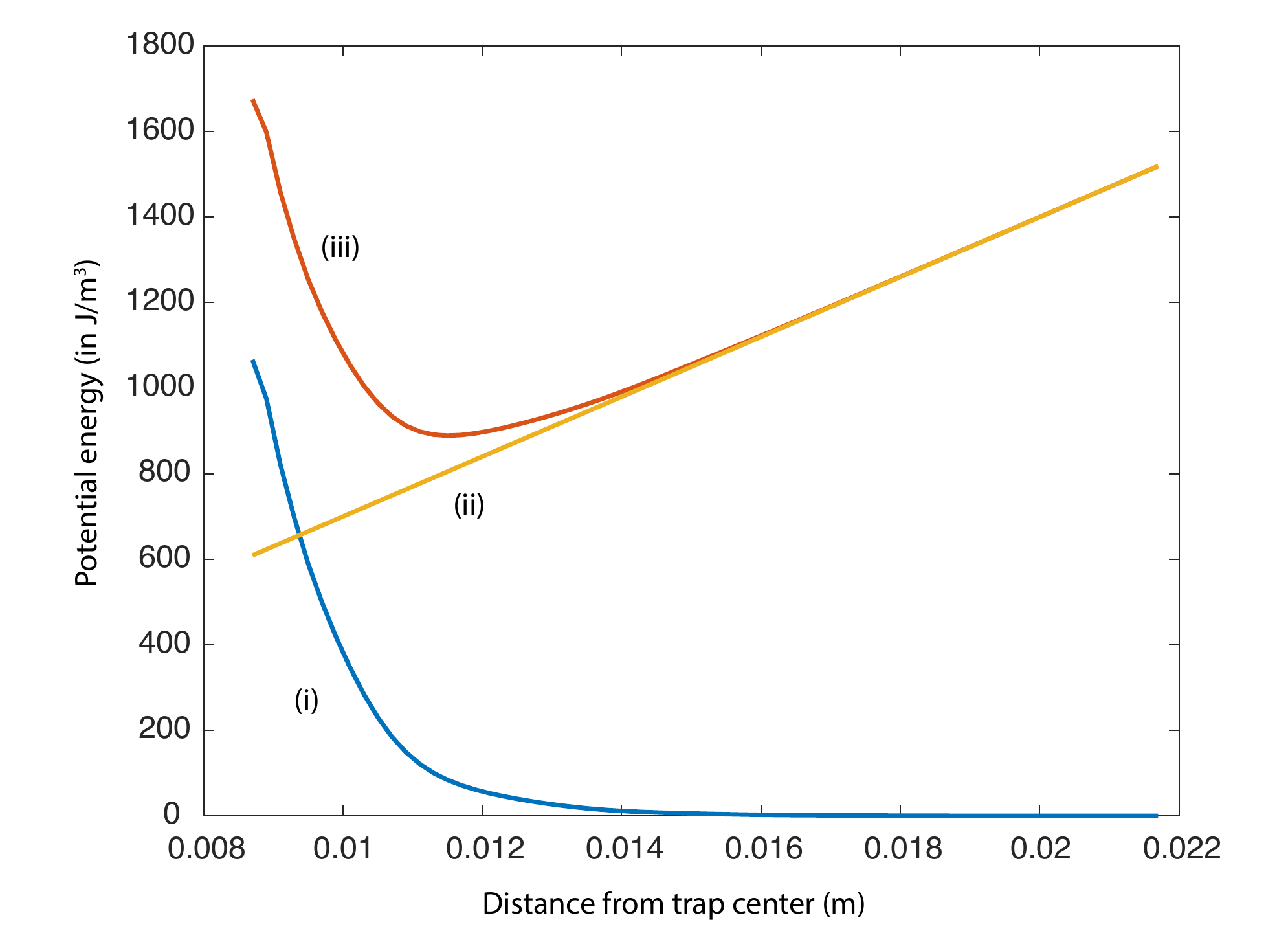}}}
\caption{Numerical simulations showing the pseudo-potential energy along $z$ as a function of the distance from the trap center (trace i), the gravitational potential (trace (ii)) and the total potential (trace (iii)).}
\label{numsim}
\end{figure}

Fig \ref{numsim}, trace (i), shows the result of a numerical simulation of $\Psi_z$, normalized to the particle volume $V$ as a function of the distance from the trap center. 
$\nabla_z F(z)$ was estimated by first fitting each of the numerically-obtained static magnetic potentials along $y$ at each position $z$. Fig.1b) in the main text, shows one such numerical simulation at a position $z=10$ mm. The modulus of the magnetic field curvature is the same along $x$. This procedure enables to extract $B''_z(z')$. The calculation is followed by a numerical integration of $B''_z(z')$ along $z'$, from $-\frac{h}{2}-z$ to $-\frac{h}{2}+z$,  and a subsequent derivation with respect to $z$. 
The parameters used in fig \ref{numsim} are $l=10$ mm,  $h=4$~mm, $\Omega/2\pi=80$~Hz and $B_{\rm sat}= 1$~T. Trace iii) is the gravitational potential $\rho g z$.  Trace (iii) shows the sum of trace (i) and (ii), showing a potential minimum at $d\approx 11.5$ mm. 

\section{Magnetic Paul trap with current carrying loops}

\begin{figure}[ht!]
\centerline{\scalebox{0.9}{\includegraphics{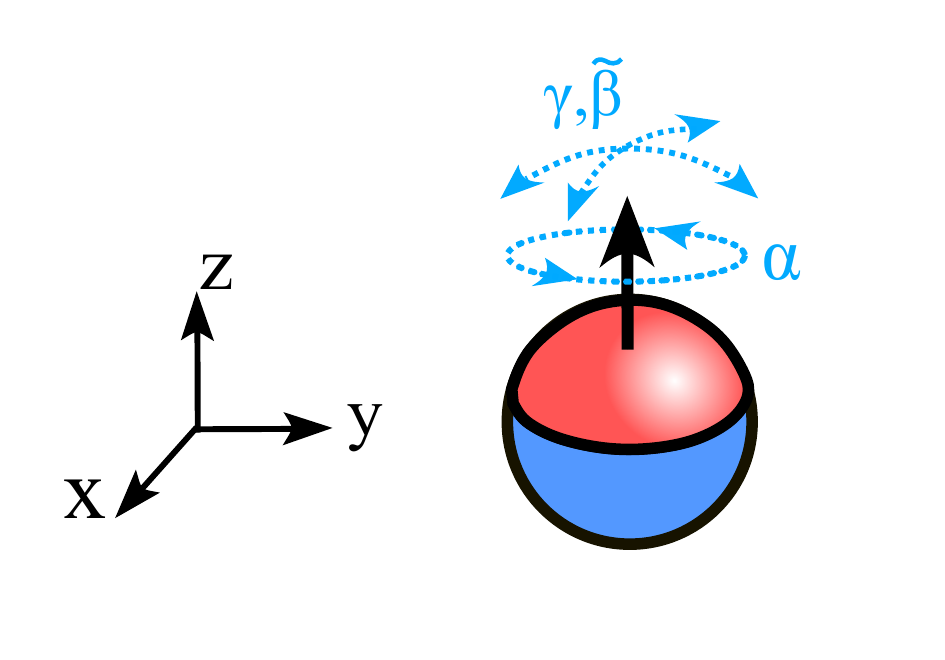}}}
\caption{Schematics showing the parametrization of both the center of mass and angular degrees of freedom.}
\label{numsim}
\end{figure}

In this section, we provide more details about the proposed current carrying loop design. 
We give the equations of motion, in the limit of small displacement, taking into account a magnetic field gradient to compensate the effect of gravity. We verify that the coupling between the angular and the center of mass (CoM) modes induced by the magnetic field gradient term is  negligible.

\subsection{System and parametrization}

We designate by $O\mathbf{e}_1\mathbf{e}_2\mathbf{e}_3$ the body-fixed reference frame of the magnet and by $O\mathbf{e}_x\mathbf{e}_y\mathbf{e}_z$ the laboratory frame. We use the Euler angle $\mathbf{u}=\,^t(\alpha,\beta,\gamma)$ in the $\bm{zyz}$ convention to parametrize the angular motion of the magnet such that $\,^t(\mathbf{e}_1,\mathbf{e}_2,\mathbf{e}_3)=\,^t R(\mathbf{u})\,^t(\mathbf{e}_x,\mathbf{e}_y,\mathbf{e}_z)$ with:

\begin{align}
R(\mathbf{u})=R_z(\alpha)R_y(\beta)R_z(\gamma)=\begin{pmatrix}
c_\alpha  & - s_\alpha  & 0\\
s_\alpha &c_\alpha & 0 \\
0 & 0 & 1 
\end{pmatrix}
\begin{pmatrix}
c_\beta  & 0  & s_\beta\\
0 & 1 & 0 \\
-s_\beta & 0 & c_\beta 
\end{pmatrix}
\begin{pmatrix}
c_\gamma  & - s_\gamma  & 0\\
s_\gamma &c_\gamma & 0 \\
0 & 0 & 1 
\end{pmatrix}.
\end{align}

where $c_\nu=\cos{(\nu)}$ and $s_\nu=\sin{(\nu)}$. Deriving the product, we get:
\begin{align}
R(\mathbf{u})=\begin{pmatrix}
c_\alpha c_\beta c_\gamma-s_\alpha s_\gamma & - s_\alpha c_\gamma-c_\alpha c_\beta s_\gamma & c_\alpha s_\beta\\
c_\alpha s_\gamma+s_\alpha c_\beta c_\gamma &c_\alpha c_\gamma-s_\alpha c_\beta s_\gamma & s_\alpha s_\beta \\
- s_\beta c_\gamma& s_\beta s_\gamma & c_\beta 
\end{pmatrix}.
\end{align}

The magnetic momentum of the ferromagnet is supposed to be attached to the particle. We consider that the magnetic momentum is oriented along the $-\mathbf{e}_1$ axis such that $\bm{\mu}=-\mu \mathbf{e}_1$ with $\mu>0$. In the laboratory-fixed coordinate system, we have:
\begin{align}
\bm{\mu}=-\mu \left((c_\alpha c_\beta c_\gamma-s_\alpha s_\gamma) \mathbf{e}_x + (c_\alpha s_\gamma+s_\alpha c_\beta c_\gamma)\mathbf{e}_y- s_\beta c_\gamma\mathbf{e}_z \right)
\end{align}
We perform the angular change of variable $\beta=\Tilde{\beta}+\pi/2$ and obtain
\begin{align}
\bm{\mu}=-\mu \left((-c_\alpha s_{\Tilde{\beta}} c_\gamma-s_\alpha s_\gamma) \mathbf{e}_x + (c_\alpha s_\gamma- s_\alpha s_{\Tilde{\beta}} c_\gamma)\mathbf{e}_y- c_{\Tilde{\beta}} c_\gamma\mathbf{e}_z \right)
\end{align}
We consider a total magnetic field $\mathbf{B}_{\rm{tot}}(\mathbf{r},t)$ composed of three different magnetic fields, a constant field $\mathbf{B}_0$, an oscillating harmonic magnetic field $\mathbf{B}_1(\bm r,t)$ and a constant magnetic field gradient $\mathbf{B}_2$ which reads: 
\begin{align}
\mathbf{B}_0&=B_0\mathbf{e}_z,\\
\begin{split}
    \mathbf{B}_1(\mathbf{r},t)&=  \frac{B_1''}{2}\cos{(\Omega t)}\left( z^2-\frac{x^2+y^2}{2}\right)\mathbf{e}_z\\ &-\frac{B_1''}{2}\cos{(\Omega t)}\left(xz\mathbf{e}_x+yz\mathbf{e}_y\right),
\end{split}\\
\mathbf{B}_2&=B_2'\left(z\mathbf{e}_z-x/2\mathbf{e}_x-y/2\mathbf{e}_y\right).
\end{align}

\subsection{Hamiltonian of the system and equation of motion}

The total Hamiltonian reads:
\begin{align}
    H=\frac{\mathbf{p}^2}{2m}+\frac{\mathbf{L}^2}{2I} -\bm{\mu}\cdot \mathbf{B}_{\rm tot}(\mathbf{r},t)+mgz,
\end{align}
where $\mathbf{p}=p_x \mathbf{e}_x + p_y \mathbf{e}_y +p_z \mathbf{e}_z$ is the CoM momentum, $\mathbf{L}=L_x \mathbf{e}_x + L_y \mathbf{e}_y +L_z \mathbf{e}_z$ is the angular momentum. We have
\begin{align}
\frac{\mathbf{L}^2}{2I}=\frac{(p_\alpha + p_\gamma\sin{\Tilde{\beta}})^2}{2I\cos{\Tilde{\beta}}^2} + 
\frac{p_{\Tilde{\beta}}^2}{2I} +  \frac{p_\gamma^2}{2I}.
\end{align}

The angular momenta $p_\alpha, p_\beta$ and $p_\gamma$ are linked to the angles by the relations:

\begin{align}
    p_{\Tilde{\beta}}&=I\dot{\Tilde{\beta}},\\
    p_{\alpha}&=I(\dot{\alpha}-\dot{\gamma} \sin{\Tilde{\beta}}),\\
    p_{\gamma}&=I(\dot{\gamma}-\dot{\alpha} \sin{\Tilde{\beta}}).    
\end{align}



Finally, we get the equations of motion:
\begin{align}
    \frac{dp_x}{dt}&= \bm{\mu}\cdot \frac{\partial \mathbf{B}}{\partial x},\\
    \frac{dp_y}{dt}&=\bm{\mu}\cdot\frac{\partial \mathbf{B}}{\partial y},\\
    \frac{dp_z}{dt}&=-mg+\bm{\mu}\cdot\frac{\partial \mathbf{B}}{\partial z},\\
        \frac{dp_\alpha}{dt}&=\frac{\partial \bm{\mu}}{\partial \alpha} \cdot \mathbf{B},\\
    \frac{dp_{\Tilde{\beta}}}{dt}&=\frac{1}{I\cos{\Tilde{\beta}}^3} (p_{\alpha} \sin{\Tilde{\beta}}+p_{\gamma})(p_{\alpha} +p_{\gamma}\sin{\Tilde{\beta}})+\frac{\partial \bm{\mu}}{\partial \Tilde{\beta}} \cdot \mathbf{B},\\
    \frac{dp_\gamma}{dt}&=\frac{\partial \bm{\mu}}{\partial \gamma} \cdot \mathbf{B}.
\end{align}

We used these equations to simulate the motion of the magnet.

\subsection{Derivation of the equation in the small motion limit}

We can calculate the forces and torques to first order in the motion variables $x,y,z,{\Tilde{\beta}},\gamma$ and we obtain:
\begin{align}
    F_x&= -\mu \frac{B_1''}{2}\cos{(\Omega t)}x-\mu \frac{B_2'}{2} (c_\alpha {\Tilde{\beta}}+s_\alpha \gamma),\\
    F_y&= -\mu \frac{B_1''}{2}\cos{(\Omega t)}y-\mu \frac{B_2'}{2} (s_\alpha \Tilde{\beta} - c_\alpha \gamma),\\
    F_z&=-mg + \mu B_2' + \mu B_1'' \cos{(\omega t)}z,\\
    \Gamma_\alpha&=0,\\
    \Gamma_{\Tilde{\beta}}&=-\mu B_0 \Tilde{\beta}-\mu \frac{B_2'}{2}\left(c_\alpha x + s_\alpha y \right),\\
    \Gamma_{\gamma}&=-\mu B_0 \gamma-\mu \frac{B_2'}{2}\left(s_\alpha x - c_\alpha y \right).
\end{align}

In order to compensate the displacement along the axis $\mathbf{e}_z$ due to the gravity, we use a field gradient which is defined by the equality $\mu B_2'=mg$. This condition does not depend on the size of the particle since both the magnetic momentum and the mass depends on the volume of the magnet. We obtain:
\begin{align}
    B_2'=\frac{\mu_0 \rho_m g}{B_{\rm{sat}}}.
\end{align}
We take $\mu_0=4\pi \times 10^{-7}~{\rm T.m.A}^{-1}$, $\rho_m=7.0 \times 10^3~  {\rm kg.m}^{-3}$, $g=9.8~  {\rm m.s}^{-2}$ and $B_{\rm{sat}} = 1.0 ~ \rm{T}$ which gives the condition:
\begin{align}
    B_2' = 8.6 \times 10^{-2}~ \rm{T}.\rm{m}^{-1}.
\end{align}

\subsection{Time-averaged equations}

Let us now derive the secular motion.
We introduce the $q$ factors:
\begin{align}
q_z=-2q_x=-2q_y=\frac{2}{\Omega^2}\frac{B_1'' B_{\rm{sat}}}{\mu_0\rho_m}.
\end{align}

Under the condition $q \leq 0.4$, we can average out the time-depending terms using the secular approximation and we obtain:

\begin{align}
    F_x/m&= - {{\Tilde{\omega}}_x}^2 x-\sqrt{\frac{I}{m}}\omega_{c}^2 (c_\alpha {\Tilde{\beta}}+s_\alpha \gamma),\\
    F_y/m&= - {{\Tilde{\omega}}_y}^2 y-\sqrt{\frac{I}{m}}\omega_{c}^2(s_\alpha \Tilde{\beta} - c_\alpha \gamma),\\
    F_z/m&=-{{\Tilde{\omega}}_z}^2 z,\\
    \Gamma_\alpha/I&=0,\\
    \Gamma_{\Tilde{\beta}}/I&=-{\omega_{\Tilde{\beta}}}^2 \Tilde{\beta}-\sqrt{\frac{m}{I}}\omega_{\rm c}^2\left(c_\alpha x + s_\alpha y \right),\\
    \Gamma_{\gamma}/I&=-{\omega_{\gamma}}^2 \gamma-\sqrt{\frac{m}{I}}\omega_{\rm c}^2\left(s_\alpha x - c_\alpha y \right).
\end{align}

with the characteristic frequencies:

\begin{align}
\omega_{\Tilde{\beta}}&=\omega_\gamma=\sqrt{\frac{5}{2} \frac{B_0 B_{\rm{sat}}}{\mu_0 \rho_m a^2}},\\
\Tilde{\omega}_z/2&=\Tilde{\omega}_x=\Tilde{\omega}_y=\Omega \frac{|q_x|}{\sqrt{2}},\\
\omega_c&=\sqrt{\sqrt{\frac{5}{2}}\frac{B_2' B_{\rm{sat}}}{\mu_0 \rho_m a}}.
\end{align}

\subsection{Numerical value and elimination of the angular-CoM coupling terms}

Experimentally, we propose to use a magnetic field $B_0 = 10 ~ {\rm mT}$ and a curvature $B_1''=10^5 ~ {\rm T.m}^{-2}$. We fixe $q_x=0.1$ which leads to the Paul trap frequency $\Omega=(2\pi)~5.0 \times 10^2 ~ {\rm Hz}$. Under these values, we obtain the typical frequencies:

 \begin{align}
 \Tilde{\omega}_x&=\Tilde{\omega}_y= (2\pi)~3.5 \times 10^1 ~{\rm Hz},\\
 \omega_{\Tilde{\beta}}&=\omega_\gamma=(2\pi)~2.7 \times 10^5 ~{\rm Hz},\\
 \omega_c&=(2\pi)~6.3 \times 10^2 ~{\rm Hz}.
 \end{align}
 
We have ${\Tilde{\omega}}_x \omega_{\Tilde{\beta}} < 10~ \omega_c^2$ so we can safely neglect the coupling between the angular and the CoM modes. 

\section{Calculation of the different inductive current in the set-up}

In this section, we examine the influence of the oscillating magnetic field on both the trapping mechanism and on the levitating particle. We conclude that for the protocol proposed, there are both negligible.

\subsection{Induction current in the micro-loops}

We calculate the Eddy current generated by the loop $1$ onto the loop $2$. The magnetic flux inside the loop $1$ equals:
\begin{align}
 \Phi_{B,1}=\pi r_1^2 B_2(t)=\pi r_1^2 \frac{\mu_0 i_2}{2r_2} \cos{(\Omega t)}.
\end{align}
The circulation of the electric field reads:
\begin{align}
\oint \mathbf{E_1}.\mathbf{dl}=2\pi r_1 E
\end{align}
The Faraday's law gives:
\begin{align}
E_1=\frac{\mu_0 i_2 \Omega}{4}\frac{r_1}{r_2}\sin{(\Omega t)}
\end{align}

Finally, we obtain the Eddy current value normalized by the initial current in the loop:
\begin{align}
i_{2 \rightarrow 1}(t)=\frac{\mu_0 \sigma S  \Omega}{4} \frac{r_1}{r_2} i_2 \sin{(\Omega t)},
\end{align}
where $\sigma$ is the electrical conductivity of gold and $S=100 ~\mu {\rm m}^2$ is the area of a slice of the gold lithography. Using $i_1/i_2=-r_1/r_2$, we get:
\begin{align}
\frac{i_{2 \rightarrow 1}(t)}{i_1}=-\frac{\mu_0 \sigma S  \Omega}{4} \sin{(\Omega t)}
\end{align}
Using the numerical values $\Omega =(2\pi) 2.0 \times 10^3 ~ {\rm Hz}$, $\sigma=4.4 \times 10^7~ {\rm S.m}^{-1}$, we obtain that this ratio is of the order of $10^{-5}$ so we can safely neglect the Eddy current generated by a loop onto the other one.

\subsection{Induction current in the levitated ferromagnet}

Induction current in the levitated ferromagnetic sphere could induce some internal heating which could be problematic at low pressure. Technically, the levitated sphere does not feel any oscillating magnetic field at the equilibrium position $(0,0,0)$. However, one has to take into account the sphere volume and the magnetic field value inside the sphere is of the order of $B_{\rm ind} \simeq B_1'' a^2 $. The current density inside the sphere equals:
\begin{align}
j \simeq \Omega \sigma a^3 B_1''
\end{align}
The power dissipated by the Joule effect then equals:
\begin{align}
P=\Omega^2 \sigma a^9 B_1''^2
\end{align}
The dependance at the power of nine of the magnet size makes the induction current heating of the order of $10^{-28} ~\rm W$. This is not sufficient to heat the internal temperature of the magnet even at ultra high vacuum.

\bibliography{SIMPT}